\documentclass[prd,aps,showpacs,,superscriptaddress]{revtex4}
\usepackage{amsmath,amssymb}  
\usepackage{bm}               
\usepackage{graphicx}         
\usepackage{epsfig}           
\newcommand{\ben}{\begin{eqnarray}}
\newcommand{\een}{\end{eqnarray}}
\begin{document}
 
\title{Transverse momentum broadening of vector boson production
       in high energy nuclear collisions}
\author{Zhong-Bo Kang}
	\email{kangzb@iastate.edu}
\affiliation{Department of Physics and Astronomy,
             Iowa State University, Ames, IA 50011, U.S.A.} 
\author{Jian-Wei Qiu}
	\email{jwq@iastate.edu}
\affiliation{Department of Physics and Astronomy,
             Iowa State University, Ames, IA 50011, U.S.A.}

\begin{abstract}
We calculate in perturbative QCD the transverse momentum broadening
of vector boson production in high energy nuclear collisions.  
We evaluate the effect of initial-state parton multiple scattering 
for the production of the Drell-Yan virtual photon and $W/Z$ bosons.
We calculate both the initial- and final-state multiple scattering
effect for the production of heavy quarkonia and 
their transverse momentum broadening in
both NRQCD and Color Evaporation model of quarkonium formation. 
We find that J/$\psi$ and $\Upsilon$ broadening
in hadron-nucleus collision is close to $2\,C_A/C_F$ times 
the corresponding Drell-Yan broadening, which gives a good 
description of existing Fermilab data.
Our calculations are also consistent with RHIC data on J/$\psi$ 
broadening in relativistic heavy ion collisions.  
We predict the transverse momentum broadening of vector boson 
(J/$\psi$, $\Upsilon$, and $W/Z$) production in relativistic
heavy ion collisions at the LHC, and discuss the role of the vector
boson broadening in diagnosing medium properties. 
\end{abstract}
\date{\today}
\maketitle

\section{Introduction}

The Relativistic Heavy Ion Collider (RHIC) at 
Brookhaven National Laboratory has been providing experimenters 
with colliding beams of heavy nuclei at 
relativistic energies as high as 100 GeV per nucleon. 
Early RHIC data from central gold-gold
collisions strongly indicate the creation of dense and hot QCD 
matter with very unusual and interesting properties 
\cite{rhic-findings}.
Characterized by its opacity to jets and other evidences, 
the new QCD matter shows an unprecedented energy density well
above the critical value predicted by lattice QCD for
establishing quark-gluon plasma (QGP), a weakly coupled gas
of quarks and gluons \cite{lattice}.  
On the other hand, the data from 
the observation of an unexpectedly large flow indicate that 
the hot QCD matter interacts strongly with itself and 
behaves like an almost ideal liquid with low shear viscosity
rather than a gas \cite{rhic-findings}.
In this paper, we investigate the role of transverse momentum 
broadening of heavy vector boson production in 
identifying the formation of new QCD matter and in 
diagnosing medium properties.

Hadronic production of a heavy vector boson, such as a virtual 
photon with a large invariant mass, a heavy quarkonium, and a $Z$ 
(or a $W$) boson requires a short-distance hard partonic collision 
with a large momentum transfer because of the large vector boson mass, 
and is dominated by the scattering of two partons from two
incoming hadron beams (or from a beam and a target).
The short-distance creation of the vector boson or the heavy quark
pair could be perturbatively calculated in Quantum Chromodynamics
(QCD), while the hadronization of the quark pair to a physical
quarkonium offers unique
perspective into the formation of QCD bound states. 
Heavy quarkonium production in high energy nuclear collisions
is of special interest in diagnosing the properties of nuclear medium
because of the two-scale nature of the creation process. 
The creation of the heavy quark pair takes place at such a
short time that it is unlikely to interfere with the dynamics of
nuclear medium which is effectively frozen.  On the other hand, 
the hadronization from the heavy quark pair to a bound quarkonium 
could be very sensitive to the medium properties.  
A change in production rate and momentum spectrum
for heavy quarkonium production from a proton-proton collision to a
nuclear collision signals a change of characteristics of the nuclear
matter and provides the opportunities to probe the matter's 
properties \cite{MS,qwg-review}. 
The challenge is to quantitatively extract interesting properties
of the nuclear matter from the change.  

In high energy hadron-nucleus and nucleus-nucleus collisions, the
interaction between the produced heavy quark pair and the nuclear 
medium (referred to as a final-state effect), as well as possible
scattering between the incoming colliding partons and the nuclear
matter before the hard collision to produce the pair (an initial-state
effect), can both change the heavy quarkonium's production rate and 
its momentum spectrum.  The effect of the final-state interaction
depends on the hadronization mechanism - how a produced heavy
quark pair becomes a bound quarkonium.
A color singlet heavy quark pair with a small color
dipole moment is less likely to interact with the nuclear medium than
a pair in a color octet configuration \cite{Qiu:1998rz,Nayak:2007mb}.  
That is, the final-state effect could be an ideal probe for exploring 
the nonperturbative formation mechanism 
in heavy quarkonium production.  But, the observed nuclear dependence 
is sensitive to not only the final-state effect but also the
initial-state effect as well as the quantum interference between
them.  Due to the local hard collision that produces the
heavy quark pair, the quantum interference between the final-state
and initial-state partonic rescattering could be suppressed by the
hard collision scale.  To understand the effect of the initial-state
interaction, we also study nuclear dependence in the production of
the Drell-Yan virtual photon as well as $Z$ (or $W$) vector boson.  
These vector bosons, if reconstructed from their leptonic decays, 
do not interact strongly once produced at the short-distance.
The initial-state effect itself can be a good probe for the dynamics
of partonic rescattering when a fast parton passes through the
nuclear matter.   

Partonic multiple rescattering in a nuclear medium before as well as
after the hard collision can modify the distribution of the vector
boson's transverse momentum $q_T$.  Each rescattering is
likely to change the momentum spectrum by an order of the typical 
transverse momentum of the partons inside the nuclear matter, which is 
significantly softer than the momentum exchange in the hard
collision.  Therefore, the change to the transverse momentum spectrum,
$d\sigma/dq_T^2$, should be most significant when $q_T$ is relatively
small.  But, the rescattering effect on the low $q_T$ spectrum is
unlikely to be calculable in perturbative QCD
\cite{Qiu-Sterman,Luo:1992fz,Luo:1993ui,Guo:1999wy}.    
On the other hand, an averaged transverse momentum square of the
produced heavy vector boson, 
\ben
\langle q_T^2\rangle
\equiv 
\int dq_T^2 \, q_T^2\,
     \frac{d\sigma_{hh\to V^*}}{dq_T^2}
\left/
\int dq_T^2 \, 
     \frac{d\sigma_{hh\to V^*}}{dq_T^2}
\right. \, ,
\label{avg-qt2}
\een
is much more inclusive and perturbatively calculable if the $q_T$ 
is integrated over a wide range 
\cite{Qiu-Sterman,Luo:1993ui,Guo:1999wy,Collins:2007nk}.  
The accumulative change from the rescattering to the averaged
transverse momentum square - transverse momentum broadening,  
$\Delta\langle q_T^2\rangle \equiv
\langle q_T^2\rangle|_{AB} - \langle q_T^2\rangle|_{hh}$,
defined as a difference between the calculable 
transverse momentum square in nuclear collision and that
in hadron-hadron collision, should be calculable too
\cite{Qiu-Sterman,Luo:1993ui}.  

The Drell-Yan transverse momentum broadening
$\Delta\langle q_T^2\rangle_{\rm DY}$ 
has been studied in perturbative QCD by
evaluating coherent partonic rescattering diagrams between 
incoming (anti)quark and the cold nuclear matter
\cite{Bodwin:1988fs,xiaofeng,Fries:2002mu}.
The rescattering leads to the broadening of the Drell-Yan 
dilepton's transverse momentum distribution.  
The broadening was shown to be proportional to the
target size or to have the $A^{1/3}$-type nuclear dependence.
The calculated nuclear dependence was found to be consistent with 
both Fermilab and CERN data \cite{xiaofeng}.  
On the other hand, as shown in Ref.~\cite{jcpeng}, 
there have been difficulties in understanding the same broadening 
in the production of heavy quarkonia (J/$\psi$ and $\Upsilon$). 
At the leading order of perturbative calculation, 
the Drell-Yan dilepton production is dominated by a quark-antiquark 
annihilation subprocess, while the heavy quarkonium production in 
hadronic collision is dominated by a gluon-gluon fusion subprocess.  
If one neglected final-state interaction in both Drell-Yan and
heavy quarkonium production, one would expect that the ratio of
broadening of heavy quarkonium over Drell-Yan is close to the 
ratio of the multiple scattering effect of a gluon over that of 
a quark (or an antiquark) inside a nuclear medium \cite{Leitch}.  
At the lowest order, 
the ratio is approximately equal to the ratio of color factors
of the lowest order gluon and (anti)quark rescattering, 
$C_A/C_F=9/4$, with $C_A=N_c=3$ and $C_F=(N_c^2-1)/(2N_c)=4/3$. 
Although the data on heavy quarkonium broadening in hadron-nucleus   
collisions shows the expected $A^{1/3}$-type nuclear dependence, 
the ratio to the Drell-Yan broadening could be as large as 5, 
twice of the naive expectation \cite{jcpeng}.  

Recently, Johnson {\it et al.} published in 
Ref.~\cite{Johnson:2006wi} a new analysis of Drell-Yan broadening, 
$\Delta \langle q_T^2\rangle_{DY}$, based on the ratio of Drell-Yan 
transverse momentum distribution on heavy nuclear targets over 
deuterium target, $R_{A/D}$.  By using the chain relation, 
$R_{A/D}=R_{A/Be}\, R_{Be/D}$, the new data from Fermilab E866 
experiment on the ratio $R_{A/Be}$ with $A=Fe,W$ over beryllium 
($Be$), and a theoretical formula for $R_{Be/D}$, 
Johnson {\it et al.} showed the result of their analysis 
in Fig.~5 of Ref.~\cite{Johnson:2006wi} 
and concluded that the observed broadening is about twice 
as large as the one published previously \cite{jcpeng},  
which implies that the heavy quarkonium broadening 
is just a factor of $C_A/C_F$ larger than the Drell-Yan 
broadening and is consistent with the naive expectation.  
The factor of two enhancement of the Drell-Yan broadening represents
a difference of more than three standard deviations between the 
central values of two different analyses 
in terms of the error size of previous analysis
or a difference of two standard deviations if one uses the much larger 
error of the new analysis. It will be interesting to see if this
new analysis changes the broadening of J/$\psi$ and $\Upsilon$ as well.  
If the ratio of the heavy quarkonium broadening over Drell-Yan 
broadening is indeed consistent with the naive expectation, we would 
have to conclude based on the discussion on final-state interaction 
in this paper that the heavy quark pair in heavy quarkonium production 
is mainly produced in a color singlet state, and the color of an octet 
pair will have to be neutralized at a very short distance.  This 
conclusion would have a very important impact on the physics of 
J/$\psi$ suppression in nuclear medium.

Since a heavy quarkonium is unlikely to be formed at the same time
when the heavy quark pair was produced \cite{Brodsky:1988xz}, 
the final-state interaction between the heavy quark pair and 
the nuclear medium could generate additional broadening.  
Since the broadening is evaluated from the transverse momentum
distribution of observed heavy quarkonia and normalized by the total
production rate, as defined in Eq.~(\ref{avg-qt2}), 
we will not discuss the impact of final-state interaction 
on the suppression of heavy quarkonium production in this paper
\cite{qwg-review,Qiu:1998rz}.  Since the final-state
interaction is sensitive to the non-perturbative formation mechanism
of the bound state, we calculate the broadening of heavy quarkonium
production in both non-relativistic QCD (NRQCD) model 
\cite{bbl-nrqcd} and Color Evaporation model \cite{CEM}.
The often used Color Singlet model \cite{Singlet} can be
interpreted as a special case of the NRQCD model. 
In the NRQCD model, the non-perturbative dynamics for a heavy quark
pair to form a bound quarkonium is organized through matrix elements
of operators that are characterized by an expansion in the relative
velocity of the pair and the pair's rotational and color quantum
numbers.  On the other hand, in the Color Evaporation model, all heavy
quark pairs with invariant mass less than the mass threshold of
producing a pair of open flavor heavy mesons 
have the same probability to become a bound quarkonium regardless the
pair's rotational and color quantum numbers.  
Rescattering of the heavy quark pair in nuclear medium 
could change the pair's rotational and color quantum numbers.
Therefore, these two models could lead to different predictions 
for the nuclear dependence of heavy quarkonium production 
in nuclear collisions.  If the difference is significant, 
an accurate measurement of the 
nuclear dependence could provide important information on the
hadronization.  We find that the heavy quarkonium broadening 
calculated in these two models have different analytical
expressions.  But, numerically, these two models predict a very
similar result for the broadening of J/$\psi$ and $\Upsilon$ in
hadron-nucleus collision.  The calculated broadening is close to
$2C_A/C_F$, which is consistent with Fermilab data
\cite{jcpeng,fix}.

We also investigate the nuclear dependence of the averaged transverse
momentum square in relativistic heavy ion collisions.  If all soft
gluons of heavy ion beams are stopped to form the hot dense medium, 
the final-state interaction between the slowly expanding or almost
stationary medium and the fast moving heavy quark pair (or a
quarkonium) of transverse momentum $q_T$ is unlikely to broaden the
$q_T$ spectrum.  Instead, the final-state interaction is likely to
slow down the heavy quarks due to the energy loss
\cite{hq-energyloss}, and could reduce the averaged transverse
momentum $\langle q_T^2 \rangle$ \cite{Kang:2007bn}.    
On the other hand, the initial-state interaction should behave 
very similar to that in hadron-nucleus collision up to a possible
difference in the medium density.  
In order to separate the effect of final-state interaction from that
of initial-state interaction, and independently study the
initial-state interaction and extract the medium density, we
calculate the broadening of $Z$ (as well as $W$) bosons 
in relativistic heavy ion collisions at the Large Hadron Collider 
(LHC).  Since there is effectively no final-state interaction 
for the production of $Z$ (or $W$) bosons 
extracted from their leptonic decay, the broadening 
is an ideal probe of the density of nuclear medium in the early stage
of the collision \cite{CERN-Yellow}.  If one would be able to
reconstruct $Z$ (or $W$) bosons from their hadronic decay, the
final-state interaction between the decaying jets and the medium could
lead to an apparent mass shift for the vector boson and provide
additional tools to extract the medium's properties. 

The rest of our paper is organized as follows. In Sec.~\ref{drellyan},
we review the perturbative QCD calculation for the transverse momentum
broadening in Drell-Yan production to set up the notation and
terminology.  In Sec.~\ref{quarkonium}, we derive the transverse
momentum broadening of heavy quarkonium production in hadron-nucleus
collision.  We calculate both initial-state and final-state multiple
scattering.  We evaluate the transverse momentum broadening
in both NRQCD model and Color Evaporation model.  We also discuss the
broadening of heavy quarkonium production in nucleus-nucleus
collisions.  In Sec.~\ref{WZ}, we calculate the broadening of
$Z$ (as well as $W$) boson production in both hadron-nucleus and
nucleus-nucleus collisions at the LHC. We argue that the
transverse momentum broadening of $Z$ (or $W$) bosons that are
reconstructed from their leptonic decay channels is an excellent
probe for initial-state rescattering and the medium density at an
early stage of relativistic heavy ion collisions.
We present our numerical study of the transverse momentum broadening
of vector boson production in Sec.~\ref{num}.  We discuss the
extrapolation of the non-perturbative matrix elements from the fixed 
target energies to collider energies.  We compare our
calculations with data from both fixed-target experiments at Fermilab, 
and collider experiments at RHIC.  We also predict the broadening at
the LHC energy.  Finally, in Sec.~\ref{summary},
we conclude with a summary that suggests directions for future work.


\section{Transverse momentum broadening in Drell-Yan production}
\label{drellyan}

The Drell-Yan production of a massive pair of leptons in hadronic
collision has been a potent probe of short-distance dynamics
in strong and electroweak interactions.  The lack of interaction
between the produced leptons and the hadronic matter makes
the Drell-Yan massive dilepton production in hadron-nucleus and 
nucleus-nucleus collisions an ideal hard probe of initial-state
partonic scattering in nuclear matter.
At the fixed target energies, the Drell-Yan process is dominated by
the production of a virtual photon of invariant mass $Q$ that 
decays into the measured lepton pairs, while the production is
dominated by the $Z$ boson at the LHC energies.  
In nuclear collisions, it is very likely that the energetic incoming
parton can have several scatterings with soft partons inside the
nuclear matter before the hard collision to produce the vector
boson.  Such initial-state multiple scattering could induce 
more soft radiation from the incoming parton and broaden 
the parton's transverse momentum.  The additional parton transverse
momentum at the hard collision leads to the broadening of the observed  
vector bosons.  

The transverse momentum distribution of the Drell-Yan vector boson
production at high $q_T$ is calculable in perturbative QCD 
\cite{CSS-fac,Berger:2001wr}.  But, the distribution 
at low $q_T\ll Q$, which is sensitive to the soft
rescattering, is not perturbatively calculable unless the resummation
of large Sudakov logarithms dominates the shape of the low $q_T$
spectrum \cite{Guo:1999wy,Collins:1984kg,Qiu:2000ga,Armesto:2007dt}.   
Each soft rescattering can only change the incoming parton's
transverse momentum by the amount close to typical virtuality of
partons inside the nuclear matter, and is too soft to be
perturbatively calculable.  Furthermore, the potential interference
between the rescattering and the parton shower, which is responsible
for the large contribution of Sudakov logarithms at low $q_T$, could
complicate the resummation of the logarithms and lead to even less
control on the low $q_T$ spectrum.  On the other hand, the 
averaged transverse momentum square, $\langle q_T^2\rangle$ 
defined in Eq.~(\ref{avg-qt2}), is much more inclusive.  
If we integrate over all kinematically allowed $q_T$, 
the $\langle q_T^2\rangle$ depends on only one single hard scale, $Q$,
the mass of the vector boson and is perturbatively calculable. The
large logarithmic contribution to the $q_T^2$-distribution from the
power of $\ln(Q^2/q_T^2)$ is suppressed by the $q_T^2$ weight.  

The transverse momentum broadening,
$\Delta\langle q_T^2\rangle \equiv
\langle q_T^2\rangle|_{AB} - \langle q_T^2\rangle|_{hh}$,
which sums over the accumulative effect of many soft rescattering,
is expressed in terms of the difference of two inclusive and 
perturbatively calculable quantities, and is therefore calculable 
in perturbative QCD \cite{Luo:1993ui,Qiu-Sterman}.
The broadening of the Drell-Yan production of lepton pairs in
hadron-nucleus collisions was first studied in terms of a
non-relativisitic QED model in Ref.~\cite{Bodwin:1988fs}.  It was
shown that initial-state interactions lead to an increase in the
average of the Drell-Yan dilepton's transverse momentum square and 
the increase is proportional to the length of the nuclear target. 
The Drell-Yan transverse momentum broadening 
was also systematically studied in terms of perturbative QCD 
collinear factorization approach in a covariant gauge \cite{xiaofeng} 
and was further studied in Ref.~\cite{Fries:2002mu} in a light-cone 
gauge.  Since we calculate the transverse momentum broadening of the
heavy quarkonium production in a covariant gauge in this paper, we
briefly review the perturbative QCD collinear factorization approach
and the covariant gauge derivation of the Drell-Yan broadening in the
rest of this section.  

\begin{figure}[hbt]
\centering
\psfig{file=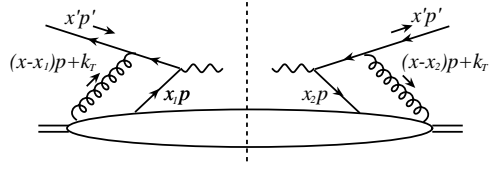,width=2.5in}
\caption{Lowest order double scattering Feynman diagram that
  contributes to the broadening of Drell-Yan transverse momentum
  distribution, which shows an antiquark of momentum $x'p'$ of
  incoming hadron scatters off a gluon of a nucleus (the bottom blob)
  before it annihilates a quark to produce a vector boson.}  
\label{dy-d}
\end{figure}

The cross section for the Drell-Yan process in hadron-nucleus
collisions, $h(p')+A(p)\to \gamma^*(q)[\to l^+l^-]+X$, where $q, p',
p$ are the four momentum of the virtual photon, the incoming hadron,
and the nucleus (per nucleon) with atomic weight $A$, respectively,
can be expanded in terms of contributions with different number of
rescattering, 
\ben
\sigma_{hA} = \sigma_{hA}^{S} + \sigma_{hA}^{D} + \dots
\label{xsec} 
\een
with superscript $S$ for single scattering, $D$ for double
scattering, and etc.  A single hard scattering is localized in
space and time, and is unlikely to provide the target length (or the
$A^{1/3}$-type nuclear size) enhancement to the cross section,
although it can get a weaker nuclear dependence to the cross section
from nuclear parton distributions \cite{Qiu-Sterman}.  
The leading contribution to the broadening of the
dilepton's transverse momentum square comes from the double
scattering \cite{xiaofeng}, 
\ben
\Delta\langle q_T^2\rangle_{\rm DY}
\approx \int dq_T^2 \, q_T^2\,
\frac{d\sigma^D_{hA}}{dQ^2dq_T^2}\left/\frac{d\sigma_{hA}}{dQ^2}
\right.  \, ,
\label{double}
\een
with the inclusive Drell-Yan cross section given by 
\ben
\frac{d\sigma_{hA}}{dQ^2}
\approx 
\frac{d\sigma_{hA}^S}{dQ^2}
\approx A \sum_q 
\int dx' \, \phi_{\bar{q}/h}(x')
\int dx \, \phi_{q/A}(x) \,
\frac{d\hat{\sigma}_{q\bar{q}}}{dQ^2}\, ,
\label{lo}
\een
where $A$ is the atomic weight of the nucleus, 
$\sum_q$ runs over all quark and antiquark flavors, 
$\phi_{\bar{q}/h}$ and $\phi_{q/A}$ represent the hadron and 
nuclear partonic distribution functions, respectively, and 
$d\hat{\sigma}_{q\bar{q}}/dQ^2$ is the lowest partonic 
$q\bar{q}$ annihilation cross section to a lepton pair of
invariant mass $Q$.  
In Eq.~(\ref{lo}) and the rest of this paper, we suppress all 
dependence on the factorization and renormalization scales. 
In Fig.~\ref{dy-d}, we sketch the leading order Feynman diagram that
contributes to the double scattering cross section, $d\sigma_{hA}^D$. 
As shown in Fig.~\ref{dy-d}, an antiquark of momentum $x'p'$ from the
incoming hadron scatters off a gluon from the nucleus (indicated by
the bottom blob) before it annihilates with a quark from the nucleus
to form a vector boson of large invariant mass, $Q$, 
which then decays into a lepton pair.
The interference diagrams, that have both gluons in the same side of
the final-state cut (the dashed line), do not contribute to the
broadening in a covariant gauge calculation \cite{xiaofeng}, while
they are very important in the light-cone gauge calculation
\cite{Fries:2002mu}.  
It is clear from the diagram that the momentum of the observed vector
boson is only sensitive to the total momentum from the nucleus, which
is equal to a sum of the gluon and quark momentum.  Therefore, the
gluon (or quark) momentum in the scattering amplitude (the left 
of the dashed line) is not necessary to be equal to the gluon (or
quark) momentum on the right of the final-state cut.  This is a 
consequence of the fact that there could be an arbitrary momentum flow 
from the nucleus through the quark line, the internal antiquark line,
and back to the nucleus from the gluon line without changing both
initial- and final-state.  To drive the double scattering contribution
to the cross section, we need to integrate over this loop momentum
for both the amplitude and complex conjugate of the amplitude, or
equivalently, the momentum flows through those two gluons in
Fig.~\ref{dy-d}.  The internal antiquark propagator following
the gluon rescattering can be very large if the gluon momentum is very
soft, and it can actually diverge if the gluon momentum vanishes.
But, it is easy to verify that the singularity of the internal
antiquark propagator when gluon momentum vanishes is not pinched.  The
integration of the gluon momentum can be deformed far away from the
on-shell singularity into a perturbative off-shell region at the order
of the hard scale $Q$; and the net result from the integration 
is given by the residue of the pole of the antiquark propagator 
\cite{Qiu-Sterman,Luo:1992fz}.  

Following the derivation in Ref.~\cite{xiaofeng}, the contribution
from the double scattering diagram in Fig.~\ref{dy-d} to the
$q_T^2$-moment of Drell-Yan cross section can be expressed as
\ben
\int dq_T^2\, q_T^2\, \frac{d\sigma_{hA}^D}{dQ^2 dq_T^2}
&=&
\sum_q\int dq_T^2 \, q_T^2 \int dx' \phi_{\bar{q}/h}(x')
\int dx\, dx_1\, dx_2\, d^2k_T\, \overline{T}_{Aq}(x,x_1,x_2,k_T,p)
\nonumber\\ 
&\times&
\overline{H}(x,x_1,x_2,k_T,p,q,x'p')\, 
\delta(q_T^2-k_T^2)\, ,
\label{double-fac}
\een
where the matrix element $\overline{T}_{Aq}$ is given by the bottom
blob in Fig.~\ref{dy-d}, which includes the propagators of all quarks 
and gluons connecting to the blob,  
\ben
\overline{T}_{Aq}(x,x_1,x_2,k_T,p)
=
\int\frac{dy^-}{2\pi}\, \frac{dy_1^-}{2\pi}\,
\frac{dy_2^-}{2\pi}\,
\int\frac{d^2y_T}{(2\pi)^2}\, 
e^{ix_1p^+y_1^-}e^{i(x-x_1)p^+y^-}e^{-i(x-x_2)p^+y_2^-}
e^{-ik_T\cdot y_T}
\nonumber\\ 
\times 
\frac{1}{2}\langle p_A|
A^+(y_2^-,0_T)\bar{\psi}(0)\gamma^+\psi(y_1^-)A^+(y^-,y_T)
|p_A\rangle \, ,
\label{Tbar}
\een 
where the subscript ``$Aq$'' indicates that the matrix element
is made of the gluon and quark field operators. 
The partonic part $\overline{H}$
in Eq.~(\ref{double-fac})  
is given by the top partonic part of the diagram in Fig.~\ref{dy-d}
with two antiquark lines traced with $(\gamma\cdot p')/2$, 
two quark lines from the nucleus traced with $(\gamma\cdot p)/2$, and
the Lorentz indices of two gluon lines from the nucleus 
contracted by $p^\alpha p^\beta$ \cite{xiaofeng}. 

\begin{figure}
\centering
\psfig{file=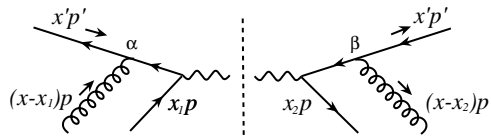,width=2.5in}
\caption{Lowest-order double scattering diagram that leads to
  the factorized partonic part, $H$, in Eq.~(\ref{H}).}
\label{dy-d0}
\end{figure}

The separation of the partonic part $\overline{H}$ from the hadronic
matrix element $\overline{T}_{Aq}$ in Eq.~(\ref{double-fac}) is not
yet a gauge invariant factorization.  The matrix element
$\overline{T}_{Aq}$ in Eq.~(\ref{Tbar}) has an explicit dependence on
the gluon field  operator $A^+$, which is not gauge covariant.  That
is, the matrix element $\overline{T}_{Aq}$ cannot be made gauge
invariant by the insertion 
of ordered gauge links between the field operators \cite{QS-fac}. 
To achieve a gauge invariant factorization, it is necessary to convert
the gluon field operator $A^+$ in the matrix element into
corresponding gluon field strength, $F^{+\alpha}$, with a transversely
polarized Lorentz index $\alpha$.  We can achieve this conversion in a
covariant gauge as follows.  We first expand the $k_T$ in the partonic
part, $\overline{H}$, around $k_T^2=0$, because $k_T^2\ll Q^2$, and
keep the first nonvanishing term, 
$H(x,x_1,x_2,p,q,x'p') = \overline{H}(x,x_1,x_2,k_T=0,p,q,x'p')$, 
which is given by the diagram in Fig.~\ref{dy-d0}.  
We then write the transverse momentum square, $q_T^2$ in
Eq.~(\ref{double-fac}) as $k_T^2$ by taking advantage of the
$\delta(q_T^2-k_T^2)$, and convert  
$k_T^2\,A^+(y_2^-,0_T)A^+(y^-,y_T)$ to
$F_\alpha^{~+}(y_2^-,0_T)F^{+\alpha}(y^-,y_T)$ by a partial
integration \cite{xiaofeng}.  Finally, upto the power corrections in
$\langle k_T^2\rangle/Q^2$, we can rewrite the $q_T^2$-moment in
Eq.~(\ref{double-fac}) as 
\ben
\int dq_T^2\, q_T^2\, \frac{d\sigma_{hA}^D}{dQ^2 dq_T^2}
=\sum_q \int dx'\, \phi_{\bar{q}/h}(x')
\int dx\, dx_1\, dx_2\, T_{Fq}^{(I)}(x,x_1,x_2,p)\, 
H(x,x_1,x_2,p,q,x'p')
\label{dy-fac}
\een 
where $T_{Fq}$ is a twist-4 parton correlation function 
defined as 
\ben
T_{Fq}^{(I)}(x,x_1,x_2,p)=
\int\frac{dy^-}{2\pi}\, \frac{dy_1^-}{2\pi}\, \frac{dy_2^-}{2\pi}\, 
e^{ix_1p^+y_1^-}e^{i(x-x_1)p^+y^-}e^{-i(x-x_2)p^+y_2^-}
\nonumber\\  
\times \frac{1}{2}\langle p_A|
F_\alpha^{~+}(y_2^-)\bar{\psi}(0)\gamma^+\psi(y_1^-)F^{+\alpha}(y^-)
|p_A\rangle   \, ,
\label{TF}
\een
with the superscript ``$(I)$'' indicates the matrix element
corresponding to the initial-state rescattering \cite{xiaofeng}.
The leading order contribution to the partonic hard part from the
diagram in Fig.~\ref{dy-d0} is \cite{xiaofeng}
\ben
H(x,x_1,x_2,p,q,x'p')
=\frac{8\pi^2\alpha_s}{N_c^2-1}\, C_F
\left[\frac{1}{2\pi}\frac{1}{x_1-x-i\epsilon}
\frac{1}{x_2-x+i\epsilon}\right]\,
\frac{d\hat{\sigma}_{q\bar{q}}}{dQ^2} \, ,
\label{H}
\een 
where two unpinched poles are from the two antiquark propagators in 
Fig.~\ref{dy-d0}.  Substituting Eq.~(\ref{H}) to
Eq.~(\ref{dy-fac}), and integrating over $x_1, x_2$, by taking the
residues of the unpinched poles (the leading pole approximation
\cite{Qiu-Sterman}), we obtain  
\ben
\int dq_T^2\, q_T^2\, \frac{d\sigma_{hA}^D}{dQ^2 dq_T^2}
=\sum_q \int dx'\, \phi_{\bar{q}/h}(x')
\int dx\, T_{q/A}^{(I)}(x) \,
\frac{d\hat{\sigma}_{q\bar{q}}}{dQ^2}
\left(\frac{8\pi^2\alpha_s}{N_c^2-1}C_F\right)\, , 
\een
with the measurable twist-4 quark-gluon correlation function
\cite{Qiu-Sterman,xiaofeng},
\ben
T_{q/A}^{(I)}(x) =
 \int \frac{dy^{-}}{2\pi}\, e^{ixp^{+}y^{-}}
 \int \frac{dy_1^{-}dy_{2}^{-}}{2\pi} \,
      \theta(y^{-}-y_{1}^{-})\,\theta(-y_{2}^{-}) 
\nonumber \\
\times \,
     \frac{1}{2}\,
     \langle p_{A}|F_{\alpha}^{\ +}(y_{2}^{-})\bar{\psi}_{q}(0)
                  \gamma^{+}\psi_{q}(y^{-})F^{+\alpha}(y_{1}^{-})
     |p_{A} \rangle \ ,
\label{TqA}
\een
where the superscript ``$(I)$'' again indicates 
the initial-state rescattering. 
From Eq.~(\ref{double}), we obtain the leading double scattering 
contribution to the Drell-Yan broadening \cite{xiaofeng},
\ben
\Delta\langle q_T^2\rangle_{\rm DY}
\approx 
\frac
{\sum_q 
  \int dx'\, \phi_{\bar{q}/h}(x')
  \int dx\, T_{q/A}^{(I)}(x)\,
  \frac{d\hat{\sigma}_{q\bar{q}}}{dQ^2}
  \left(\frac{8\pi^2\alpha_s}{N_c^2-1}\, C_F\right)}
{A\sum_q 
  \int dx'\, \phi_{\bar{q}/h}(x')
  \int dx\, \phi_{q/A}(x)\,
  \frac{d\hat{\sigma}_{q\bar{q}}}{dQ^2}}\, . 
\een
By using the model proposed for the twist-4 parton correlation
functions \cite{Luo:1992fz,xiaofeng}
\ben
T_{q/A}^{(I)}(x)=\lambda^2\, A^{4/3}\, \phi_{q/A}(x)\, ,
\label{ansatz}
\een
we can express the Drell-Yan broadening in a much simpler form
\cite{xiaofeng}  
\ben
\Delta\langle q_T^2\rangle_{\rm DY}
=C_F
\left(\frac{8\pi^2\alpha_s}{N_c^2-1}\, \lambda^2\, A^{1/3}\right) \, ,
\label{broaden-dy} 
\een
with an unknown non-perturbative parameter $\lambda^2$ defined in
Eq.~(\ref{ansatz}).  The leading contribution to the Drell-Yan
broadening in Eq.~(\ref{broaden-dy}) shows a clear $A^{1/3}$-type
dependence and is proportional to the color factor $C_F$ from the
rescattering between an antiquark (or a quark) and a gluon.

\section{Transverse momentum broadening in heavy quarkonium
  production} 
\label{quarkonium}

In this section we use the same technique reviewed in last section to
calculate the transverse momentum broadening of heavy quarkonium
production in both hadron-nucleus and nucleus-nucleus collisions.

The heavy quarkonium's transverse momentum broadening in
hadron-nucleus collision was often attributed to the 
initial-state multiple scattering 
between the active parton of the projectile and soft partons of the
nuclear target before the hard collision to produce the heavy quark
pair \cite{Leitch}.  Calculation of such initial-state rescattering
should be very similar to that for the Drell-Yan broadening, except
that the quark-antiquark annihilation is accompanied by a much larger 
gluon-gluon fusion subprocess.  If one considers only the gluon-gluon
fusion subprocess, one should expect to have Eq.~(\ref{broaden-dy})
for the heavy quarkonium broadening with the overall color factor
$C_F$ replaced by $C_A=N_c=3$ due to the difference in color factors
between gluon rescattering and quark rescattering.  The initial-state
rescattering alone leads to the naive expectation for the ratio of
broadening between heavy quarkonium and Drell-Yan as $C_A/C_F=2.25$,
which is much smaller than the data \cite{jcpeng}.

However, as discussed in the Introduction of this paper, the net
broadening of heavy quarkonium's transverse momentum in hadron-nucleus
collision is a combined effect of the initial-state interaction and
final-state rescattering between the produced heavy quark pair and the
nuclear matter.  This is because a heavy quarkonium is very
unlikely to form at the same time when the heavy quark pair was
produced.  Since the final-state rescattering is sensitive
to the detailed dynamics that transmutes a heavy quark pair into a
bound quarkonium, we calculate the final-state
contribution to heavy quarkonium broadening in both NRQCD and Color 
Evaporation models.

\begin{figure}[hbt]
\centering
\psfig{file=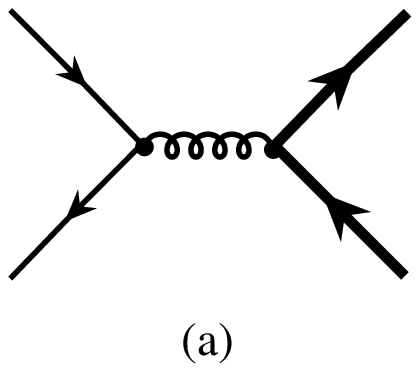,width=1.0in}
\hskip 0.5in
\psfig{file=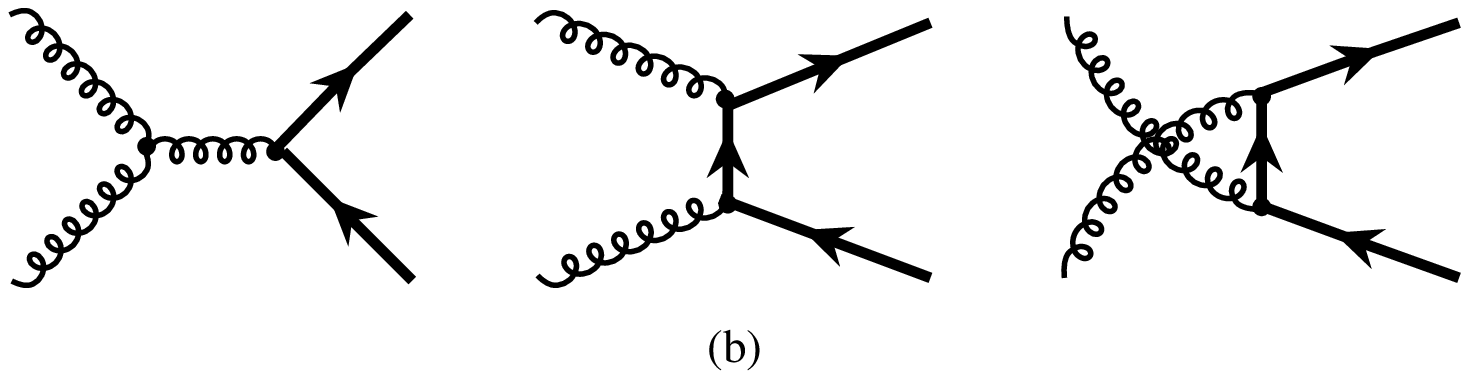,width=3.7in}
\caption{Lowest order Feynman diagram for light quark-antiquark
  annihilation (a) and for gluon-gluon fusion to a pair of heavy
  quark.}   
\label{hq-lo}
\end{figure}

\subsection{Color Evaporation Model}
\label{broadening-cem}

In the Color Evaporation model (CEM), heavy quarkonium production is
factorized into two steps: the production of a pair of
heavy quarks with an invariant mass $Q$ followed by a non-perturbative 
hadronization process with an universal transition probability for the
pair to become a bound quarkonium \cite{CEM}.  It was assumed that the
transition probability is the same for all heavy quark pairs whose
invariant mass is less than the mass threshold of producing two open
flavor heavy mesons, and the cross section for producing a heavy
quarkonium, $H$, can be written as \cite{CEM}
\ben
\sigma_{hA\to H}^{\rm CEM}
=
F_{Q\bar{Q}\to H}\int_{4m_Q^2}^{4M_Q^2} dQ^2\, 
\frac{d\sigma_{hA\to Q\bar{Q}}}{dQ^2} \, ,
\label{cem-fac}
\een
where $F_{Q\bar{Q}\to H}$ is a non-perturbative transition
probability and is independent of the color and angular momentum
of the heavy quark pair.  There is one transition probability for each
heavy quarkonium state, $H$.  In Eq.~(\ref{cem-fac}), the
inclusive cross section for producing 
a pair of heavy quarks of invariant mass $Q$ can be factorized 
as \cite{CSS-HQ}
\ben
\frac{d\sigma_{hA\to Q\bar{Q}}}{dQ^2} 
= A\, \sum_{a,b}
\int dx' \, \phi_{a/h}(x')
\int dx \, \phi_{b/A}(x) \,
\frac{d\hat{\sigma}_{ab\to Q\bar{Q}}}{dQ^2} \, ,
\label{cem-lo}
\een
where $\sum_{a,b}$ sum over all parton flavors, and
$d\hat{\sigma}_{ab\to Q\bar{Q}}/dQ^2$ is a short-distance hard part
for two partons of flavor $a$ and $b$ to produce a pair of heavy
quarks of invariant mass $Q$.  At the lowest order, they are given by
the light quark-antiquark annihilation and gluon-gluon fusion
subprocess, as sketched in Fig.~\ref{hq-lo}.
The transition probability in Eq.~(\ref{cem-fac}) is assumed to
be universal and independent of how the heavy quark pair was produced.
It fixes the overall normalization for the cross section of
heavy quarkonium production in different collision processes 
and provides the predictive power of the model.  
The model has been reasonably successful when comparing with data of
inclusive heavy quarkonium production \cite{cem-hp,cem-cdf}.  

\begin{figure}[hbt]
\centering
\psfig{file=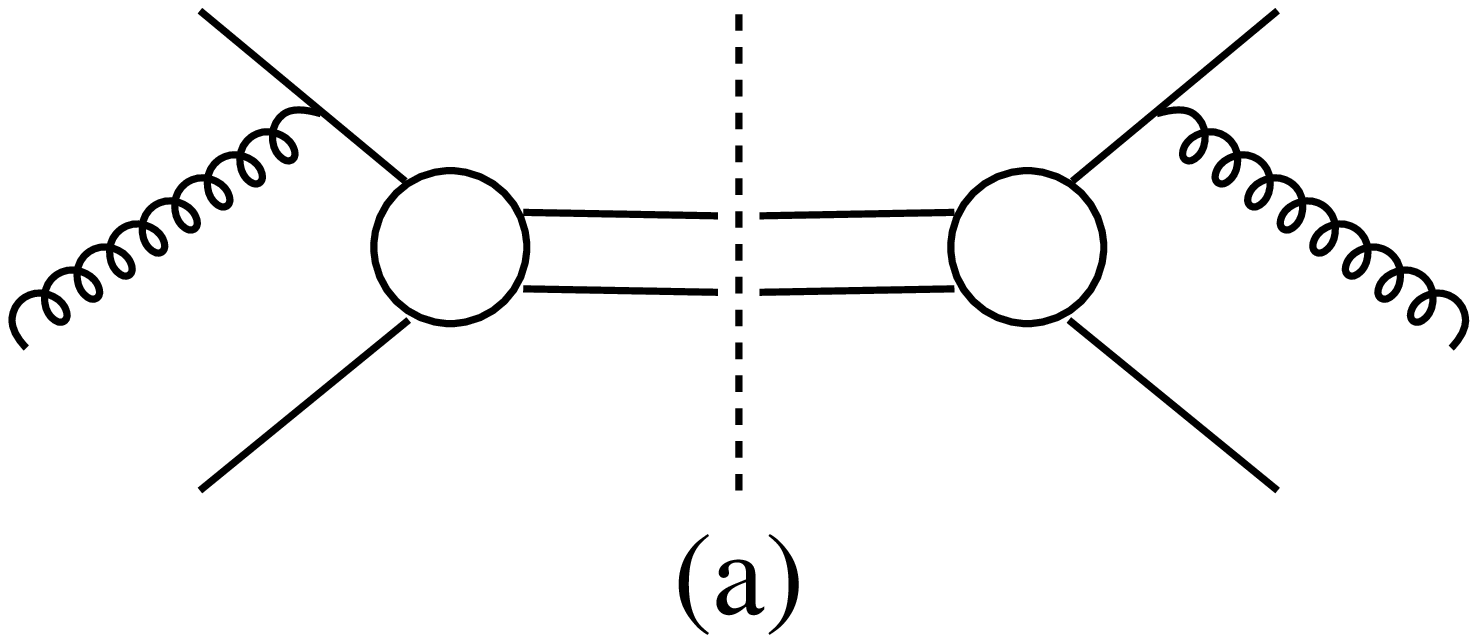,width=1.6in}\\
\vskip 0.1in
\psfig{file=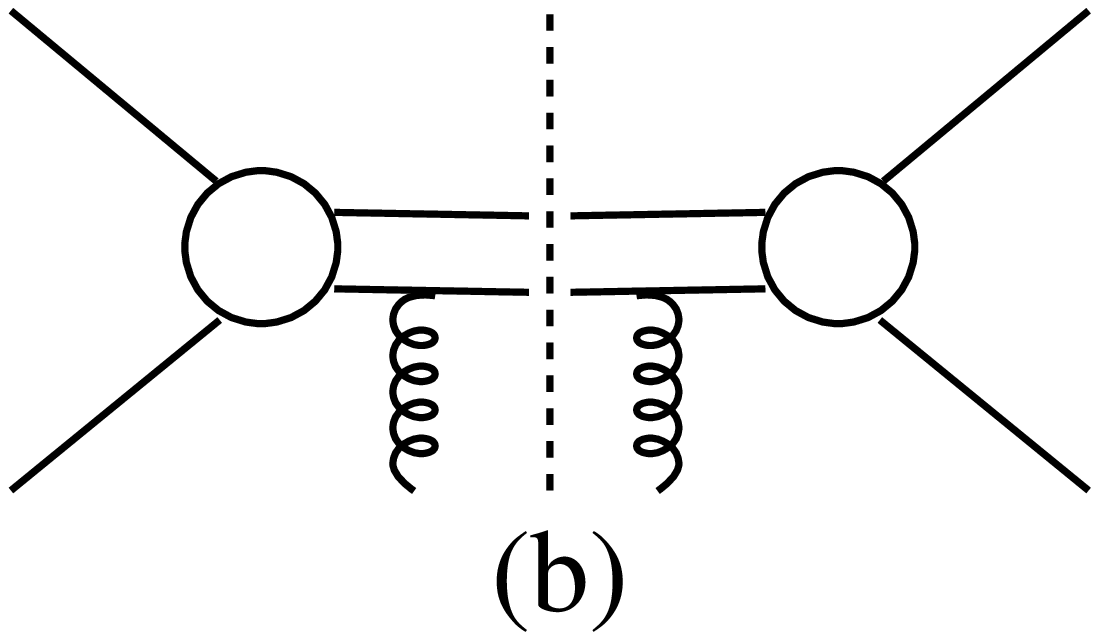,width=1.2in}
\hskip 0.1in
\psfig{file=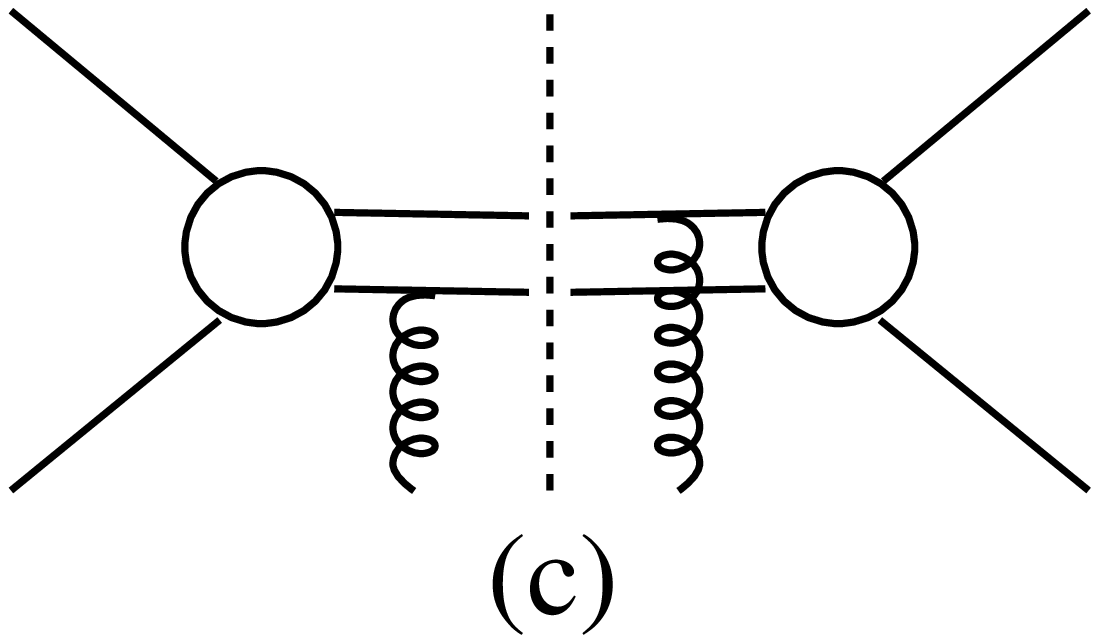,width=1.2in}
\hskip 0.1in
\psfig{file=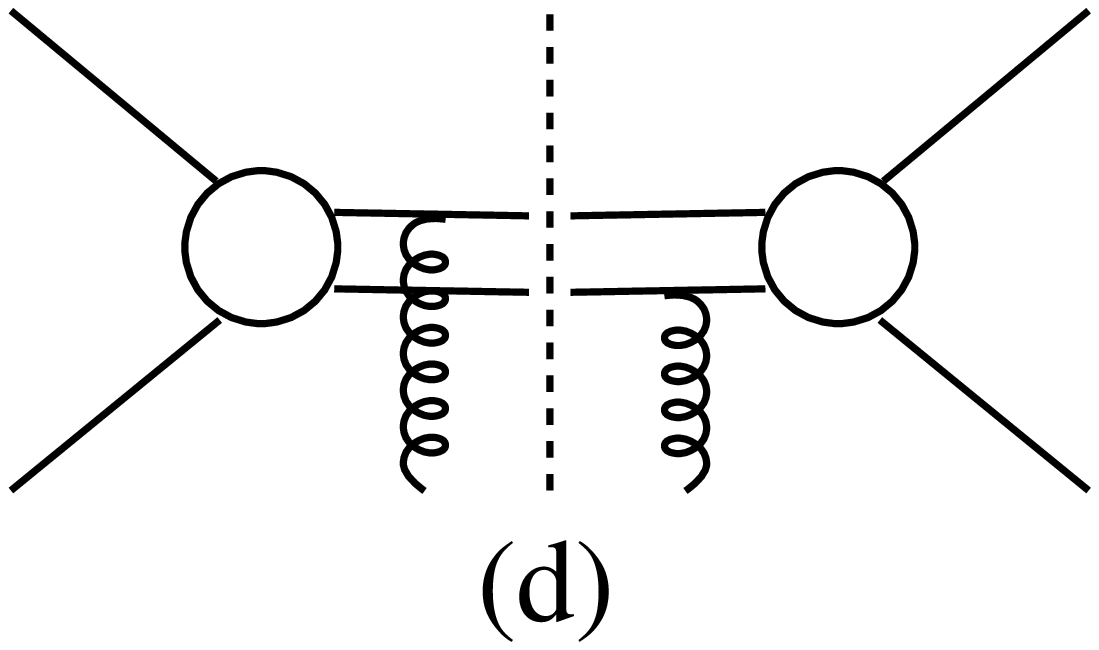,width=1.2in}
\hskip 0.1in
\psfig{file=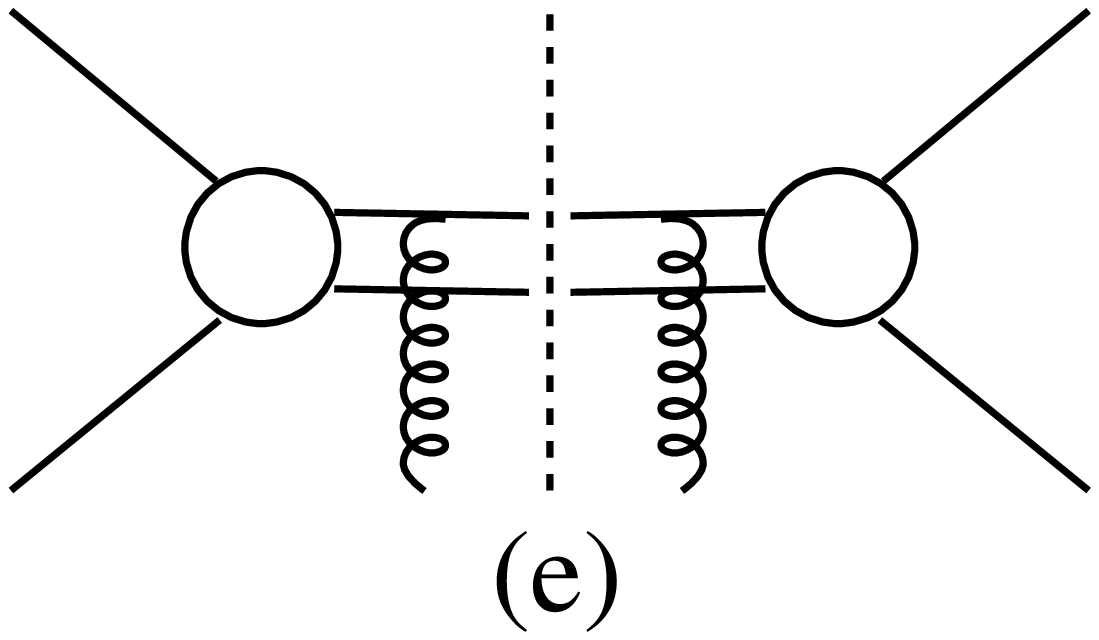,width=1.2in}
\caption{Leading order double scattering diagrams for 
  $q\bar{q}\to Q\bar{Q}$: initial-state double scattering (a), 
  and final-state double scattering (b), (c), (d), and (e).}
\label{hq-qq-d}
\end{figure}
\begin{figure}[hbt]
\centering
\psfig{file=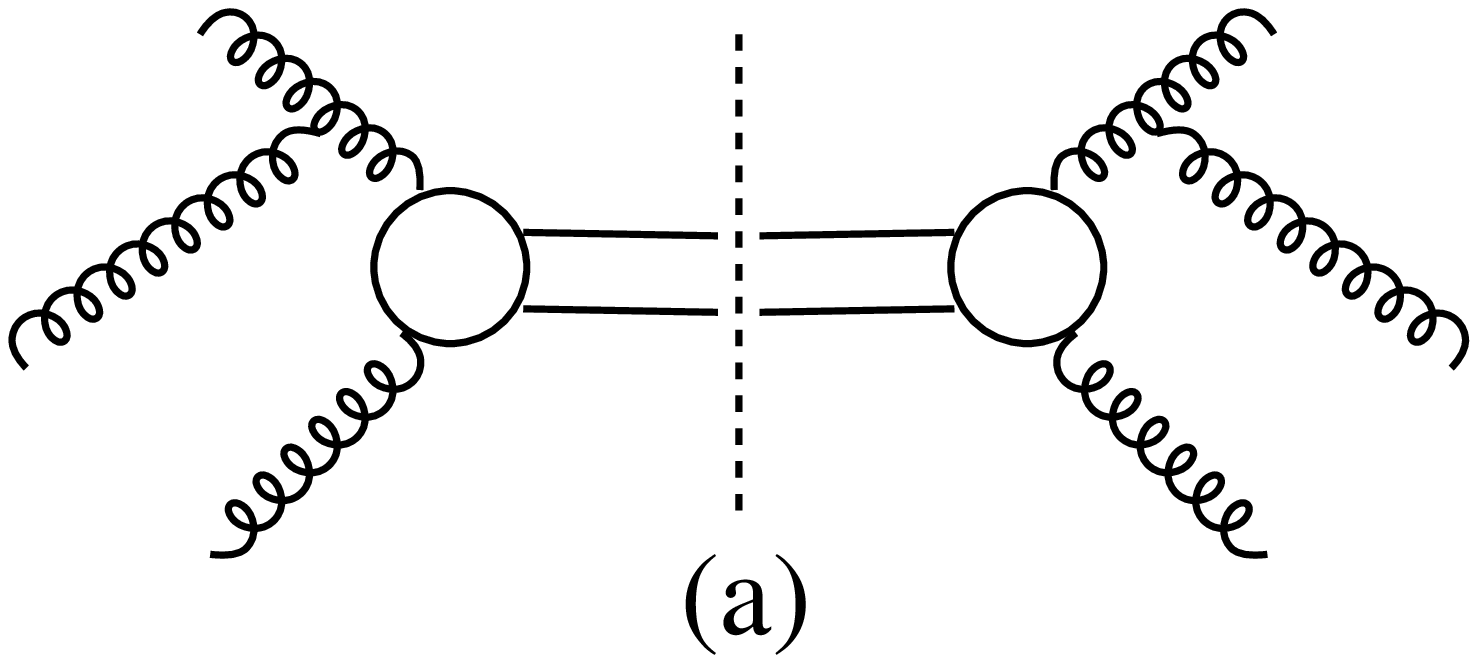,width=1.6in}\\
\vskip 0.1in
\psfig{file=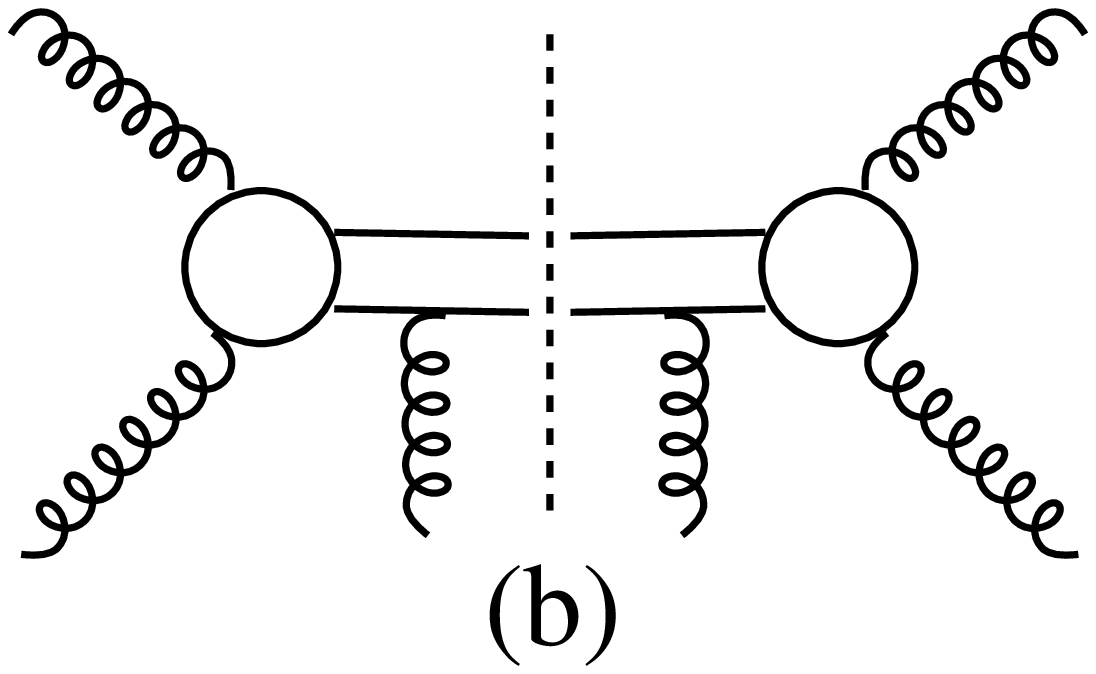,width=1.2in}
\hskip 0.1in
\psfig{file=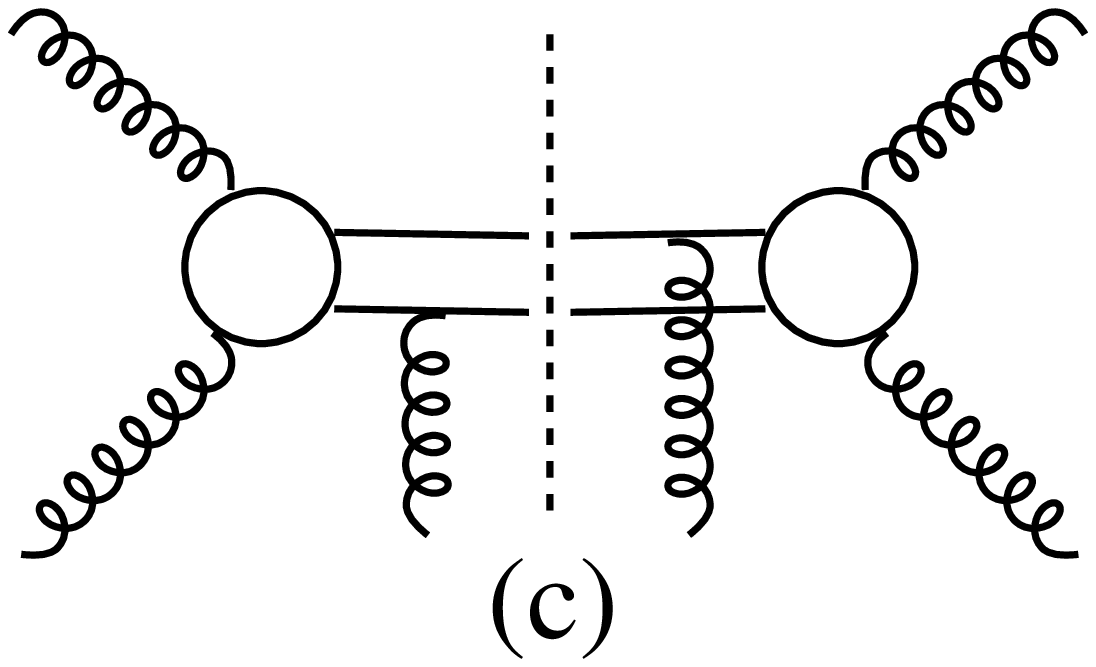,width=1.2in}
\hskip 0.1in
\psfig{file=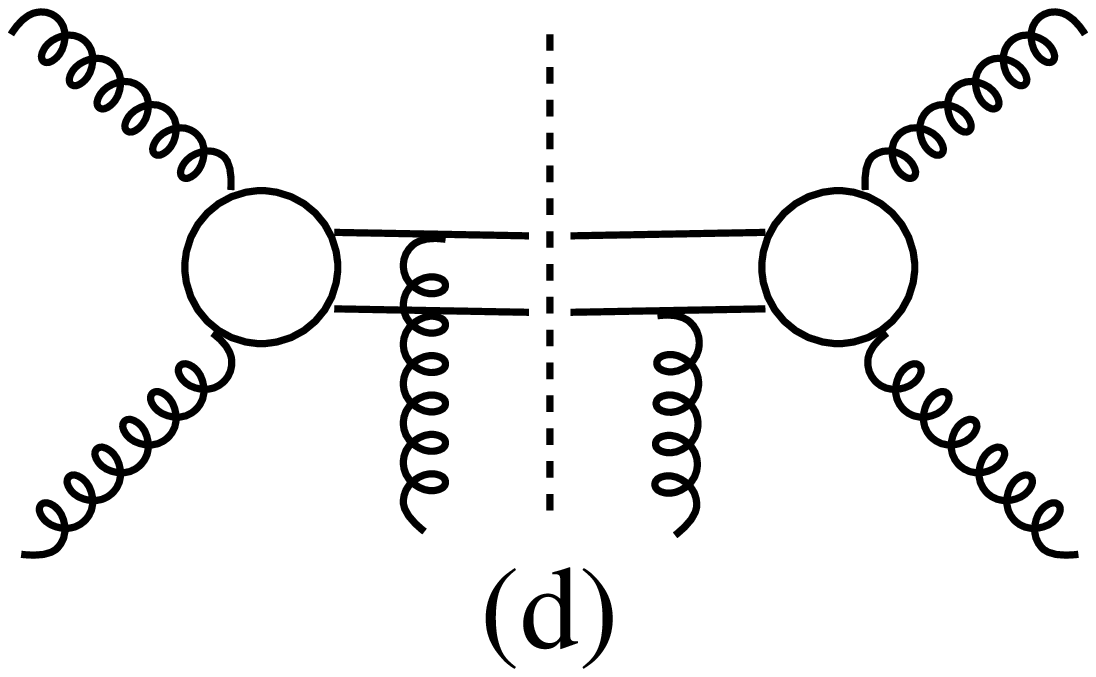,width=1.2in}
\hskip 0.1in
\psfig{file=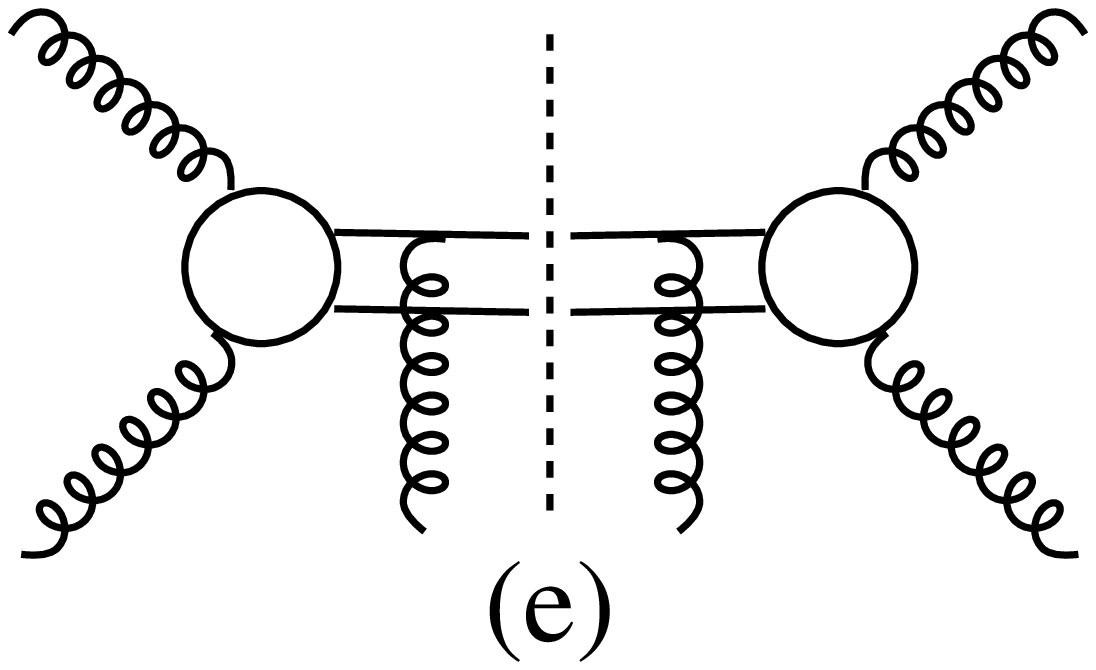,width=1.2in}
\caption{Leading order double scattering diagrams for 
  $gg\to Q\bar{Q}$: initial-state double scattering (a), 
  and final-state double scattering (b), (c), (d), and (e).}
\label{hq-gg-d}
\end{figure}

Similar to Fig.~\ref{dy-d0}, the leading order double scattering
diagrams for producing a heavy quark pair are
sketched in Fig.~\ref{hq-qq-d} for quark-antiquark annihilation
subprocess, and in Fig.~\ref{hq-gg-d} for gluon-gluon fusion
subprocess, respectively.  
The blob in the quark-antiquark annihilation subprocess
in Fig.~\ref{hq-qq-d} is given by the diagram in Fig.~\ref{hq-lo}(a), 
and the blob in the gluon-gluon fusion subprocess in
Fig.~\ref{hq-gg-d} is given by a sum of the three diagrams in
Fig.~\ref{hq-lo}(b).

In CEM, the transverse momentum broadening of a heavy
quarkonium is equal to the transverse momentum broadening of the
parent heavy quark pair, since the transition probability from a heavy
quark pair to a bound quarkonium is given by a constant,
$F_{Q\bar{Q}\to H}$.
We use the same method reviewed in last section to calculate the
transverse momentum broadening of the heavy quark pairs.
Similar to Eq.~(\ref{dy-fac}) in the Drell-Yan case, we have   
\ben
\int dq_T^2\, q_T^2\, \frac{d\sigma_{hA\to Q\bar{Q}}^D}{dQ^2 dq_T^2}
=\sum_q
 \int dx' \phi_{\bar{q}/h}(x')
 \int dx\, dx_1\, dx_2\, 
\big[ T_{Fq}^{(I)}(x,x_1,x_2,p)
      H_{q\bar{q}\to Q\bar{Q}}^{(I)}(x,x_1,x_2,p,q,x'p')
\nonumber\\ 
    + T_{Fq}^{(F)}(x,x_1,x_2,p)
      H_{q\bar{q}\to Q\bar{Q}}^{(F)}(x,x_1,x_2,p,q,x'p') \big]
\nonumber\\ 
+\int dx' \phi_{g/h}(x')
\int dx\, dx_1\, dx_2\,  
\big[ T_{FF}^{(I)}(x,x_1,x_2,p)
      H_{gg\to Q\bar{Q}}^{(I)}(x,x_1,x_2,p,q,x'p')
\nonumber\\
    + T_{FF}^{(F)}(x,x_1,x_2,p)
      H_{gg\to Q\bar{Q}}^{(F)}(x,x_1,x_2,p,q,x'p') \big]\, , 
\label{hq-fac}
\een
where the superscripts, ``$(I)$'' and ``$(F)$'', indicate the
initial-state and final-state rescattering, respectively, 
and the matrix element $T_{Fq}^{(I)}$ is given in Eq.~(\ref{TF}). 
$T_{FF}^{(I)}$ is given by
\ben
T_{FF}^{(I)}(x,x_1,x_2,p)=\int\frac{dy^-}{2\pi}
\frac{dy_1^-}{2\pi}\frac{dy_2^-}{2\pi}
e^{ix_1p^+y_1^-}e^{i(x-x_1)p^+y^-}e^{-i(x-x_2)p^+y_2^-}\nonumber\\ 
\times 
\langle p_A| F_\alpha^{~+}(y_2^-)
F^{\sigma+}(0)F^+_{~\sigma}(y_1^-)F^{+\alpha}(y^-)|p_A\rangle\, .
\label{TG}
\een
The matrix elements with final-state rescattering, $T_{Fq}^{(F)}$ 
and $T_{FF}^{(F)}$, have the same expressions as corresponding 
matrix elements with initial-state rescattering, 
since the field operators in the definition of the multiparton matrix
elements in the collinear factorization approach commute on the
light-cone \cite{QS-fac}.
 
The partonic parts, $H_{q\bar{q}\to Q\bar{Q}}^{(I,F)}$, are given by
the Feynman diagrams in Fig.~\ref{hq-qq-d} with the quark line from
hadron (top) traced with $(\gamma\cdot p')/2$, the quark line from
nucleus (bottom) traced with $(\gamma\cdot p)/2$, and gluon lines
contracted with $p^\alpha p^\beta$.  The diagram with initial-state
rescattering in Fig.~\ref{hq-qq-d}(a) contributes to $H_{q\bar{q}\to
  Q\bar{Q}}^{(I)}$ as
\ben
H_{q\bar{q}\to Q\bar{Q}}^{(I)}
=H_{q\bar{q}\to Q\bar{Q}}^{(\ref{hq-qq-d}a)}
=\frac{8\pi^2\alpha_s}{N_c^2-1}\, C_F
\left[\frac{1}{2\pi}\frac{1}{x_1-x-i\epsilon}
\frac{1}{x_2-x+i\epsilon}\right]
\frac{d\hat{\sigma}_{q\bar{q}\to Q\bar{Q}}}{dQ^2}\, ,
\label{qq-i}
\een
where the lowest order partonic cross section from $q\bar{q}$
annihilation to a heavy quark pair of invariant mass $Q$ is 
given in Ref.~\cite{Benesh:1994du}.  
All four diagrams with the final-state rescattering, in
Figs.~\ref{hq-qq-d}(b), (c), (d), and (e), contribute to 
$H_{q\bar{q}\to Q\bar{Q}}^{(F)}$ as
\ben
H_{q\bar{q}\to Q\bar{Q}}^{(F)}
=H_{q\bar{q}\to Q\bar{Q}}^{(\ref{hq-qq-d}b+\ref{hq-qq-d}c
+\ref{hq-qq-d}d+\ref{hq-qq-d}e)}
=\frac{8\pi^2\alpha_s}{N_c^2-1}\, C_A
\left[\frac{1}{2\pi}\frac{1}{x_1-x+i\epsilon}
\frac{1}{x_2-x-i\epsilon}\right]
\frac{d\hat{\sigma}_{q\bar{q}\to Q\bar{Q}}}{dQ^2}\, .
\label{qq-f}
\een 
The final-state contribution in Eq.~(\ref{qq-f}) is very similar to 
the initial-state contribution in Eq.~(\ref{qq-i}) except the overall
color factor and the location of the unpinched poles.  The difference
in the location of the pinched poles, indicated by the sign difference
of the $i\epsilon$, is a consequence of the order of the rescattering 
taken place either before or after the hard collision.  The overall
color factor for the final-state interaction, $C_A$ in
Eq.~(\ref{qq-f}), indicates that as far as the 
color is concerned, the rescattering of a heavy quark pair is
effectively the same as that of a color-octet gluon when the effect is
calculated in the Color Evaporation model. 

Similarly, we evaluate the double scattering diagrams from gluon-gluon
fusion subprocess in Fig.~\ref{hq-gg-d} and obtain their contribution
to the partonic hard part,
\ben
H_{gg\to Q\bar{Q}}^{(I)}=
H_{gg\to Q\bar{Q}}^{(\ref{hq-gg-d}a)}=
\frac{8\pi^2\alpha_s}{N_c^2-1}\, C_A
\left[\frac{1}{2\pi}\frac{1}{x_1-x-i\epsilon}
\frac{1}{x_2-x+i\epsilon}\right]
\frac{d\hat{\sigma}_{gg\to Q\bar{Q}}}{dQ^2}\, ,
\label{gg-i} 
\een
for the initial-state double scattering, and 
\ben
H_{gg\to Q\bar{Q}}^{(F)}=
H_{gg\to Q\bar{Q}}^{\ref{hq-gg-d}b+\ref{hq-gg-d}c
+\ref{hq-gg-d}d+\ref{hq-gg-d}e}=
\frac{8\pi^2\alpha_s}{N_c^2-1}\, C_A
\left[\frac{1}{2\pi}\frac{1}{x_1-x+i\epsilon}
\frac{1}{x_2-x-i\epsilon}\right]
\frac{d\hat{\sigma}_{gg\to Q\bar{Q}}}{dQ^2} \, ,
\label{gg-f}
\een
for the final-state double scattering.  We find that the contribution
to the gluon-gluon fusion subprocess
from the final-state double scattering is the same as that from the
initial-state interaction.  That is because the rescattering effect of  
a heavy quark pair is the same as that of a color-octet gluon when
the effect is calculated in the Color Evaporation model.

Substituting the partonic hard parts in Eqs.~(\ref{qq-i}),
(\ref{qq-f}), (\ref{gg-i}), and (\ref{gg-f}) into Eq.~(\ref{hq-fac}),
integrating over the momentum fractions, $x_1$ and $x_2$ of the
rescattering gluons under the leading pole approximation, 
we obtain the leading double scattering contribution to the 
$q_T^2$-moment of producing a heavy quark pair in hadron-nucleus
collisions,   
\ben
\int dq_T^2\, q_T^2\, \frac{d\sigma_{hA\to Q\bar{Q}}^D}{dQ^2 dq_T^2}
=
\left[\frac{8\pi^2\alpha_s}{N_c^2-1}\right]
\left(\sum_q
\int dx' \phi_{\bar{q}/h}(x')
\int dx\left[C_F\, T_{q/A}^{(I)}(x) + C_A\, T_{q/A}^{(F)}(x)\right] 
\frac{d\hat{\sigma}_{q\bar{q}\to Q\bar{Q}}}{dQ^2}
\right.\nonumber\\
\left.+
\int dx' \phi_{g/h}(x')
\int dx \left[C_A\, T_{g/A}^{(I)}(x) + C_A\, T_{g/A}^{(F)}(x)\right] 
\frac{d\hat{\sigma}_{gg\to Q\bar{Q}}}{dQ^2}
\right)\, ,
\label{hq-diff}
\een
where the quark-gluon correlation function, $T_{q/A}^{(I)}$, is given
in Eq.~(\ref{TqA}), the $T_{g/A}^{(I)}$ is given by 
\ben
T_{g/A}^{(I)}(x) &=&
 \int \frac{dy^{-}}{2\pi}\, e^{ixp^{+}y^{-}}
 \int \frac{dy_1^{-}dy_{2}^{-}}{2\pi} \,
      \theta(y^{-}-y_{1}^{-})\,\theta(-y_{2}^{-}) 
\nonumber \\
&\times &
\frac{1}{xp^+}\,
\langle p_A| F_\alpha^{~+}(y_2^-)
F^{\sigma+}(0)F^+_{~\sigma}(y^-)F^{+\alpha}(y_1^-)|p_A\rangle\, ,
\label{TgA}
\een
and $T_{q/A}^{(F)}$ and $T_{g/A}^{(F)}$ are given by 
the same expressions in Eq.~(\ref{TqA}) and Eq.~(\ref{TgA}),
respectively, except the $\theta$-functions are replaced
as \cite{Qiu-Sterman}
\ben
\theta(y^{-}-y_{1}^{-})\,\theta(-y_{2}^{-}) 
\to
\theta(y_{1}^{-}-y^{-})\,\theta(y_{2}^{-}) \, ,
\label{theta-ps}
\een
due to the different order of rescattering (or the sign of the
$i\epsilon$ for the unpinched poles).

By integrating over the invariant mass of the heavy quark pair 
we can define the heavy quarkonium transverse momentum broadening in
CEM as
\ben
\Delta\langle q_T^2\rangle_{\rm HQ}^{\rm CEM}
\approx \int dq_T^2 \, q_T^2
\int_{4m_Q^2}^{4M_Q^2}dQ^2 \,
\frac{d\sigma^D_{hA\to Q\bar{Q}}}{dQ^2dq_T^2}
\left/
\int_{4m_Q^2}^{4M_Q^2}dQ^2 \,
\frac{d\sigma_{hA\to Q\bar{Q}}}{dQ^2}
\right.  \, ,
\label{double-hq}
\een
where $m_Q$ and $M_Q$ are heavy quark mass and open flavor heavy meson
mass, respectively.  

As pointed out in Ref.~\cite{QS-fac}, the field operators on the
light-cone in the definition of the multiparton matrix elements, as
those in Eqs.~(\ref{TqA}) and (\ref{TgA}) commute.  
The matrix element with initial-state rescattering is equal to 
corresponding matrix element with final-state rescattering, if the
phase space interaction of these two matrix elements, such as the
$\int dy^-\, dy_1^-\, dy_2^- \theta(y^- -y_1^-)\theta(-y_2^-)$ in
Eq.~(\ref{TqA}) for initial-state rescattering and
$\int dy^-\, dy_1^-\, dy_2^- \theta(y_1^- -y^-)\theta(y_2^-)$ for 
the corresponding final-state rescattering, are the same
\cite{xiaofeng}.  However, the phase space integration for the
final-state interaction in heavy quarkonium production may not cover
the full size of the nuclear medium if the heavy quark pair becomes
a physical quarkonium or transmutes into a color singlet pre-hadron
quarkonium state before the pair exits the nuclear medium.  
Rescattering between a quarkonium and nuclear medium and that 
between a colored heavy quark pair and the same medium 
could be different, and lead to a different heavy quarkonium
broadening.  

It was argued in Ref.~\cite{Brodsky:1988xz} that a physical
quarkonium state is likely to form outside nuclear matter in
hadron-nucleus collision.  Therefore, the matrix elements with final-
and initial-state rescattering could have the same phase space
interaction.  If we assume that the matrix elements with
the final- and initial-state rescattering are the same, 
$T_{q/A}^{(F)}(x)=T_{q/A}^{(I)}(x)$ and
$T_{g/A}^{(F)}(x)=T_{g/A}^{(I)}(x)$, and assume the same model for both
twist-4 quark-gluon and gluon-gluon correlation functions, 
\ben
T_{g/A}^{(I)}(x)=\lambda^2\, A^{4/3}\,\phi_{g/A}(x)\, ,
\label{ansatz-g}
\een
we can express the heavy quarkonium broadening in hadron-nucleus
collisions as, 
\ben
\Delta\langle q_T^2\rangle_{\rm HQ}^{\rm CEM}
=\left(\frac{8\pi^2\alpha_s}{N_c^2-1}\, \lambda^2\, A^{1/3}\right) 
\frac{(C_F+C_A)\, \sigma_{q\bar{q}}+2\,C_A\, \sigma_{gg}}
     {\sigma_{q\bar{q}}+\sigma_{gg}} \, ,
\label{cem-qt2}
\een
where $\sigma_{q\bar{q}}$ and $\sigma_{gg}$ are the lowest order
inclusive cross sections from the $q\bar{q}\to Q\bar{Q} \to H$ and
$gg\to Q\bar{Q}\to H$ subprocess, respectively.  
They are given by \cite{Qiu:1998rz,Benesh:1994du}
\ben
\sigma_{q\bar{q}}
&=&F_{Q\bar{Q}\to H}
\int_{4m_Q^2}^{4M_Q^2}dQ^2
\sum_q\int dx' \phi_{\bar{q}/h}(x')
\int dx\, \phi_{q/A}(x)\, 
\frac{d\hat{\sigma}_{q\bar{q}\to Q\bar{Q}}}{dQ^2}
\\ 
\sigma_{gg}
&=&F_{Q\bar{Q}\to H}
\int_{4m_Q^2}^{4M_Q^2}dQ^2\int dx' \phi_{g/h}(x')
\int dx\, \phi_{g/A}(x)\, 
\frac{d\hat{\sigma}_{gg\to Q\bar{Q}}}{dQ^2} \, ,
\een
where the lowest order partonic cross sections to produce a pair of 
$Q\bar{Q}$ are given in Ref.~\cite{Benesh:1994du}.  
If the gluon-gluon subprocess dominates the heavy quarkonium
production rate, $\sigma_{gg}\gg \sigma_{q\bar{q}}$, we have 
\ben
\Delta\langle q_T^2\rangle_{\rm HQ}^{\rm CEM}
\approx
2\, C_A
\left(\frac{8\pi^2\alpha_s}{N_c^2-1}\, \lambda^2\, A^{1/3}\right) 
\, . 
\label{cem-qt2-gg}
\een
By comparing the Drell-Yan broadening in Eq.~(\ref{broaden-dy}) and 
the leading heavy quarkonium broadening in Eq.~(\ref{cem-qt2-gg}),
we conclude that the leading contribution to heavy quarkonium
transverse momentum broadening in hadron-nucleus collision, calculated
in CEM, is about $2\,C_A/C_F=4.5$ times Drell-Yan broadening.

\subsection{Non-Relativistic QCD Model}
\label{broadening-nrqcd}

The application of NRQCD to the production of a heavy quarkonium $H$
in hadronic collisions relies on the
proposed factorization formalism \cite{bbl-nrqcd},    
\ben
\sigma_{hA\to H}^{\rm NRQCD}
= A\,
\sum_{a,b}\int dx'\, \phi_{a/h}(x')
\int dx\, \phi_{b/A}(x)
\left[\sum_{n} H_{ab\to Q\bar{Q}[n]}
\langle {\cal O}^H(n)\rangle\right] 
\label{nrfac}
\een
where $a$ and $b$ run over all parton flavors, and 
$H_{ab\to Q\bar{Q}[n]}$ are perturbatively calculable coefficient
functions for producing the heavy $Q\bar{Q}[n]$ states.  
The state of the heavy quark pair, $[n]$, is characterized by  
the pair's rotational, $^{2s+1}L_{J}$, and color quantum numbers.  
The coefficient function for producing
each $Q\bar{Q}[n]$ state is perturbatively calculable in a power
series of the strong coupling constant, $\alpha_s$.  
The matrix elements of $\langle {\cal O}^H(n)\rangle$ in
Eq.~(\ref{nrfac}) describe the non-perturbative hadronization dynamics 
and give the probability for the pair to become a physical heavy
quarkonium $H$ \cite{bbl-nrqcd,nqs-fac}.  
The matrix elements should be universal.   
That is necessary for the predictive power of the NRQCD
formalism.  The expansion in Eq.~(\ref{nrfac}) is organized according
to the effective power of the heavy quark pair's relative velocity.  
Although it still lacks a fully compelling proof for
the NRQCD factorization formula in Eq.~(\ref{nrfac}) \cite{nqs-fac}, 
the formalism for heavy quarkonium production has had many successes, 
in particular, its success in interpreting the CDF data on J/$\psi$
and $\psi'$ production as a function of transverse momentum
\cite{qwg-review,hq-review}.  

In NRQCD model of heavy quarkonium production, the transition
probability from a heavy quark pair to a bound quarkonium is sensitive
to the pair's rotational and color quantum numbers.  Partonic multiple
scattering has a potential to change not only the heavy quark pair's 
momentum, but also the pair's color and other quantum numbers.
Therefore, the transverse momentum broadening calculated in NRQCD
model is not necessary the same as that calculated in CEM in last
subsection.  If the difference is significant, a precise measurement
of transverse momentum broadening could shed some lights on heavy
quarkonium's production mechanism. 

We assume that the $q_T^2$-moment of heavy quarkonium production, 
$\int dq_T^2 (q_T^2)^n d\sigma_{hA\to H}^{\rm NRQCD}/dq_T^2$
with $n\ge 0$, can be factorized in the same way as the 
$0^{\rm th}$-moment given in Eq.~(\ref{nrfac}).  We calculate the
leading double scattering contribution to the broadening of the first
moment.  Like the calculation done in CEM in last subsection, the
partonic double scattering diagrams are given in Figs.~\ref{hq-qq-d}
and \ref{hq-gg-d}.  Similar to Eq.~(\ref{hq-fac}), we obtain  
\ben
\int dq_T^2\, q_T^2\, \frac{d\sigma_{hA\to H}^D}{dq_T^2}
=\sum_q
 \int dx' \phi_{\bar{q}/h}(x')
 \int dx\, dx_1\, dx_2\, 
\big[ T_{Fq}^{(I)}(x,x_1,x_2,p)
      H_{q\bar{q}\to H}^{(I)}(x,x_1,x_2,p,q,x'p')
\nonumber\\ 
    + T_{Fq}^{(F)}(x,x_1,x_2,p)
      H_{q\bar{q}\to H}^{(F)}(x,x_1,x_2,p,q,x'p') \big]
\nonumber\\ 
+\int dx' \phi_{g/h}(x')
\int dx\, dx_1\, dx_2\,  
\big[ T_{FF}^{(I)}(x,x_1,x_2,p)
      H_{gg\to H}^{(I)}(x,x_1,x_2,p,q,x'p')
\nonumber\\
    + T_{FF}^{(F)}(x,x_1,x_2,p)
      H_{gg\to H}^{(F)}(x,x_1,x_2,p,q,x'p') \big]\, , 
\label{nrqcd-fac}
\een
with the partonic cross sections defined as
\ben
H_{q\bar{q}\to H}^{(I,F)}(x,x_1,x_2,p,q,x'p')
=\sum_n
H_{q\bar{q}}^{(I,F)}(n)\,
\langle {\cal O}^H(n)\rangle \, ,
\nonumber\\
H_{gg\to H}^{(I,F)}(x,x_1,x_2,p,q,x'p')
=\sum_n 
H_{gg}^{(I,F)}(n)\,
\langle {\cal O}^H(n)\rangle \, ,
\label{combine}
\een
where $\sum_n$ sums over all possible $Q\bar{Q}$ states, $n$, 
with appropriate spin and color quantum numbers \cite{Cho}, and 
$H_{q\bar{q}}^{(I,F)}(n)$ and $H_{gg}^{(I,F)}(n)$, whose  
dependence on parton momentum fractions and kinematic
variables are suppressed, represent partonic hard parts for 
producing a heavy quark pair at a quantum state $n$ from
quark-antiquark annihilation subprocess and gluon-gluon fusion
subprocess, respectively. 

The partonic parts for the quark-antiquark annihilation subprocess, 
$H_{q\bar{q}}^{(I,F)}(n)$ in Eq.~(\ref{combine}), 
are derived from diagrams in Fig.~\ref{hq-qq-d}. 
The single diagram in Fig.~\ref{hq-qq-d}(a) contributes
to $H_{q\bar{q}}^{(I)}(n)$.  Under the leading pole
approximation, the initial-state rescattering does not change the 
nature of the $s$-channel $q\bar{q}\to Q\bar{Q}$ subprocess, which
produces a heavy $Q\bar{Q}$ pair in a color octet and spin-1 state:  
$n={}^3S_1^{(8)}$.  The corresponding hard part is given by
\ben
H_{q\bar{q}}^{(I)}({}^3S_1^{(8)})=
\frac{8\pi^2\alpha_s}{N_c^2-1}\, C_F
\left[\frac{1}{2\pi}\frac{1}{x_1-x-i\epsilon}
\frac{1}{x_2-x+i\epsilon}\right]
H_{q\bar{q}}^{(0)}({}^3S_1^{(8)}) \, ,
\label{nrqq-i}
\een
where $H_{q\bar{q}}^{(0)}({}^3S_1^{(8)})$ is the lowest order
short-distance coefficient for $q\bar{q}\to Q\bar{Q}({}^3S_1^{(8)})$
subprocess and is given by
\ben
H_{q\bar{q}}^{(0)}({}^3S_1^{(8)})=
\frac{\pi^3\alpha_s^2}{M^3}\frac{16}{27}\delta(\hat{s}-M^2)\, ,
\label{qq-s1c8-lo}
\een
with the mass of a quarkonium: $M=2m_Q$.  
The final-state rescattering effect to the $q\bar{q}$ annihilation
subprocess comes from the four diagrams 
in Figs.~\ref{hq-qq-d}(b), (c), (d), and (e).  The additional gluon
rescattering in the final-state allows more quantum states for the
produced $Q\bar{Q}$ pair.  At this order, we have the following 
nonvanishing states: $n={}^3S_1^{(8)}, {}^3P_{J=1,2}^{(1,8)}$, 
and corresponding hard parts,
\ben
H_{q\bar{q}}^{(F)}({}^3S_1^{(8)})=
\frac{8\pi^2\alpha_s}{N_c^2-1}\, C_A
\left[\frac{1}{2\pi}\frac{1}{x_1-x+i\epsilon}
\frac{1}{x_2-x-i\epsilon}\right]
H_{q\bar{q}}^{(0)}({}^3S_1^{(8)})\, ,
\nonumber\\ 
H_{q\bar{q}}^{(F)}({}^3P_{J=1,2}^{(1,8)})=
\frac{8\pi^2\alpha_s}{N_c^2-1}
\left[\frac{1}{2\pi}\frac{1}{x_1-x+i\epsilon}
\frac{1}{x_2-x-i\epsilon}\right]
H_{q\bar{q}}^{(1)}({}^3P_{J=1,2}^{(1,8)})\, ,
\label{nrqq-f}
\een
where $H_{q\bar{q}}^{(0)}({}^3S_1^{(8)})$ is given in
Eq.~(\ref{qq-s1c8-lo}) and the leading order $P$-wave contribution  
is given by
\ben
H_{q\bar{q}}^{(1)}({}^3P_{1}^{(8)})=
\frac{5}{3}H_{q\bar{q}}^{(1)}({}^3P_{2}^{(8)})= 
\frac{\pi^3\alpha_s^2}{M^3}\frac{40}{27}\frac{1}{3m_Q^2}
\delta(\hat{s}-M^2)\, ,
\nonumber\\
H_{q\bar{q}}^{(1)}({}^3P_{1}^{(1)})=
\frac{5}{3}H_{q\bar{q}}^{(1)}({}^3P_{2}^{(1)})=
\frac{\pi^3\alpha_s^2}{M^3}\frac{64}{81}
\frac{1}{3m_Q^2}\delta(\hat{s}-M^2)\, .
\label{qq-H-lo}
\een
The nonvanishing contribution to the ${}^3P_{J=1,2}^{(1,8)}$ states 
is a consequence of the gluon rescattering, which effectively provides
a $gg\to Q\bar{Q}$ subprocess.   

Similarly, partonic parts for the gluon-gluon fusion subprocess, 
$H_{gg}^{(I,F)}(n)$ in Eq.~(\ref{combine}), 
are derived from diagrams in Fig.~\ref{hq-gg-d}.
Unlike the quark-antiquark annihilation subprocess, 
the heavy quark pair produced in gluon-gluon subprocess 
can have more than one quantum state.  For the initial-state
rescattering diagram in Fig.~\ref{hq-gg-d}(a), the heavy quark pair
with $n={}^1S_0^{(1,8)}, {}^3P_{J=0,2}^{(1,8)}$ can all give
nonvanishing contribution to $H_{gg}^{(I)}(n)$.  The four final-state 
rescattering diagrams in Figs.~\ref{hq-gg-d}(b), (c), (d), and (e)
can produce heavy quark pairs with 
$n={}^1S_0^{(8)}$ and ${}^3P_{J=0,2}^{(8)}$.  
We obtain the hard parts from gluon-gluon fusion diagrams in 
Fig.~\ref{hq-gg-d} as
\ben
H_{gg}^{(I)}({}^1S_0^{(1,8)}, {}^3P_{J=0,2}^{(1,8)})
&=&
\frac{8\pi^2\alpha_s}{N_c^2-1}\, C_A
\left[\frac{1}{2\pi}\frac{1}{x_1-x-i\epsilon}
\frac{1}{x_2-x+i\epsilon}\right]
H_{gg}^{(0)}({}^1S_0^{(1,8)}, {}^3P_{J=0,2}^{(1,8)})\, ,
\label{nrgg-i}\\
H_{gg}^{(F)}({}^1S_0^{(8)}, {}^3P_{J=0,2}^{(1,8)})
&=&\frac{8\pi^2\alpha_s}{N_c^2-1}\,C_A
\left[\frac{1}{2\pi}\frac{1}{x_1-x+i\epsilon}
\frac{1}{x_2-x-i\epsilon}\right]
H_{gg}^{(0)}({}^1S_0^{(8)}, {}^3P_{J=0,2}^{(8)})\, ,
\label{nrgg-f}
\een
where $H_{gg}^{(0)}({}^1S_0^{(1,8)}, {}^3P_{J=0,2}^{(1,8)})$ are the
lowest order cross sections without the rescattering,
\ben
H_{gg}^{(0)}({}^1S_0^{(1,8)})
&=&
\frac{\pi^3\alpha_s^2}{M^3}B_n\delta(\hat{s}-M^2)\, ,
\nonumber\\
H_{gg}^{(0)}({}^3P_0^{(1,8)})
&=&
\frac{\pi^3\alpha_s^2}{M^3}B_n\frac{3}{m_Q^2}\delta(\hat{s}-M^2)\, ,
\nonumber\\
H_{gg}^{(0)}({}^3P_2^{(1,8)})
&=&\frac{\pi^3\alpha_s^2}{M^3}B_n\frac{4}{5m_Q^2}
\delta(\hat{s}-M^2) \, ,
\label{nrgg-lo}
\een
with a color factor $B_n=2/9$ for a color-singlet state and 
$5/12$ for a color-octet state, respectively.

Having obtained the short-distance hard parts, we can derive the
leading double scattering contribution to the $q_T^2$-moment of heavy
quarkonium production in Eq.~(\ref{nrqcd-fac}).  In this paper, we 
limit ourselves to the {\it direct} production of spin-1, $S$-wave
heavy quarkonia, such as J/$\psi$, $\Upsilon$ and etc.  After
neglecting the color singlet $Q\bar{Q}$ states not equal to ${}^3S_1$, 
we are left with the following $Q\bar{Q}$ states, which contribute to
the direct production of a $H({}^3S_1)$ heavy quarkonium, 
\ben
q\bar{q}\to Q\bar{Q}({}^3S_1^{(8)},{}^3P_{J=1,2}^{(8)})
        \to H({}^3S_1)\, ,
\nonumber\\
gg\to Q\bar{Q}({}^1S_0^{(8)},{}^3P_{J=0,2}^{(8)})
  \to H({}^3S_1)\, .
\label{channel}
\een
From Eq.~(\ref{combine}), we obtain by summing over all $Q\bar{Q}$
states in Eq.~(\ref{channel}),
\ben
H_{q\bar{q}\to H}^{(I)}
&=&
\frac{8\pi^2\alpha_s}{N_c^2-1}
\left[\frac{1}{2\pi}\frac{1}{x_1-x-i\epsilon}
\frac{1}{x_2-x+i\epsilon}\right]
C_F \, 
H_{q\bar{q}}^{(0)}({}^3S_1^{(8)}) 
\langle {\cal O}^H({}^3S_1^{(8)})\rangle 
\nonumber \\
&=&
\frac{8\pi^2\alpha_s}{N_c^2-1}
\left[\frac{1}{2\pi}\frac{1}{x_1-x-i\epsilon}
\frac{1}{x_2-x+i\epsilon}\right]
C_F \, 
\hat{\sigma}_{q\bar{q}}^{(0)}\, ,
\label{nrqq-H-i}
\een
where the lowest order quark-antiquark annihilation cross section is
defined as 
\ben
\hat{\sigma}_{q\bar{q}}^{(0)}
\equiv 
H_{q\bar{q}}^{(0)}({}^3S_1^{(8)})\,
\langle {\cal O}^H({}^3S_1^{(8)})\rangle
=\frac{\pi^3\alpha_s^2}{M^3}\frac{16}{27}\delta(\hat{s}-M^2)\,
\langle {\cal O}^H({}^3S_1^{(8)}) \rangle
\label{sigma-qq-0}
\een
with the nonperturbative NRQCD matrix element,
$\langle {\cal O}^H({}^3S_1^{(8)}) \rangle$, for a
$Q\bar{Q}[{}^3S_1^{(8)}]$ pair to become a heavy quarkonium $H$. 
Similarly, we have the final-state rescattering contribution,
\ben
H_{q\bar{q}\to H}^{(F)}
&=&
\frac{8\pi^2\alpha_s}{N_c^2-1}
\left[\frac{1}{2\pi}\frac{1}{x_1-x+i\epsilon}
\frac{1}{x_2-x-i\epsilon}\right]
\left(
C_A\, H_{q\bar{q}}^{(0)}({}^3S_1^{(8)})
\langle {\cal O}^H({}^3S_1^{(8)})\rangle
\right. 
\nonumber \\
& & {\hskip 2.2 in} \left.
+H_{q\bar{q}}^{(1)}({}^3P_{1}^{(8)})
\langle {\cal O}^H({}^3P_1^{(8)})\rangle
+H_{q\bar{q}}^{(1)}({}^3P_{2}^{(8)})
\langle {\cal O}^H({}^3P_2^{(8)})\rangle
\right)
\nonumber \\
&=&
\frac{8\pi^2\alpha_s}{N_c^2-1}
\left[\frac{1}{2\pi}\frac{1}{x_1-x+i\epsilon}
\frac{1}{x_2-x-i\epsilon}\right]
\left(
C_A\, \hat{\sigma}_{q\bar{q}}^{(0)}
+ \hat{\sigma}_{q\bar{q}}^{(1)}
\right)\, ,
\label{nrqq-H-f}
\een
where we defined the $P$-wave contribution as
\ben
\hat{\sigma}_{q\bar{q}}^{(1)}
&\equiv &
 H_{q\bar{q}}^{(1)}({}^3P_{1}^{(8)})
\langle {\cal O}^H({}^3P_1^{(8)})\rangle
+H_{q\bar{q}}^{(1)}({}^3P_{2}^{(8)})
\langle {\cal O}^H({}^3P_2^{(8)})\rangle
\nonumber\\
&=&\frac{\pi^3\alpha_s^2}{M^3}\frac{80}{27}\,
\delta(\hat{s}-M^2)\, \langle {\cal O}^H({}^3P_0^{(8)})\rangle \, .
\label{sigma-qq-1}
\een
In deriving the last equation, we used the heavy quark spin symmetry
\cite{Cho},    
\ben
\langle {\cal O}^H({}^3P_J^{(8)})\rangle
=(2J+1)\langle {\cal O}^H({}^3P_0^{(8)})\rangle \, .
\een

From Eqs.~(\ref{nrgg-i}) and (\ref{nrgg-f}), after we neglect
the color singlet channels that have wrong spin and angular momentum
quantum numbers, the gluonic contribution from the initial-state and
final-state rescattering to the direct production of $H({}^3S_1)$ are
effectively the same except the sign of the $i\epsilon$ for the
unpinched poles,
\ben
H_{gg\to H}^{(I)} 
&=&
\frac{8\pi^2\alpha_s}{N_c^2-1}
\left[\frac{1}{2\pi}\frac{1}{x_1-x-i\epsilon}
\frac{1}{x_2-x+i\epsilon}\right]
C_A\, \hat{\sigma}_{gg}^{(0)}\, ,
\nonumber \\
H_{gg\to H}^{(F)}
&=&
\frac{8\pi^2\alpha_s}{N_c^2-1}
\left[\frac{1}{2\pi}\frac{1}{x_1-x+i\epsilon}
\frac{1}{x_2-x-i\epsilon}\right]
C_A\, \hat{\sigma}_{gg}^{(0)}\, ,
\label{nrgg-H}
\een
where the gluon-gluon fusion cross section is defined as
\ben
\hat{\sigma}_{gg}^{(0)}
\equiv
\frac{\pi^3\alpha_s^2}{M^3}\frac{5}{12}\,\delta(\hat{s}-M^2)
\left[\langle {\cal O}^H({}^1S_0^{(8)})\rangle
+\frac{7}{m_Q^2}\langle {\cal O}^H({}^3P_0^{(8)})\rangle\right]\, .
\label{sigma-gg-0}
\een
In deriving Eq.~(\ref{nrgg-H}), the heavy quark spin symmetry was used
to relate the matrix elements of different $P$-wave states to the
matrix element of $\langle {\cal O}^H({}^3P_0^{(8)})\rangle$.

Substituting the partonic cross sections in Eqs.~(\ref{nrqq-H-i}),
(\ref{nrqq-H-f}), and (\ref{nrgg-H}) into Eq.~(\ref{nrqcd-fac}), and
integrating over gluon momentum fractions, $x_1$ and $x_2$, by taking
the residue, we obtain the double scattering contribution to the
$q_T^2$-moment of heavy quarkonium production in NRQCD model as,
\ben
\int dq_T^2\, q_T^2\, \frac{d\sigma_{hA\to H}^D}{dq_T^2}
=
\left[\frac{8\pi^2\alpha_s}{N_c^2-1}\right]
\Bigg(\sum_q
 \int dx' \phi_{\bar{q}/h}(x')
 \int dx\, \left[
  T_{q/A}^{(I)}(x)\, C_F\, \hat{\sigma}_{q\bar{q}}^{(0)}
+ T_{q/A}^{(F)}(x)\, \left( C_A\, \hat{\sigma}_{q\bar{q}}^{(0)}
                          + \hat{\sigma}_{q\bar{q}}^{(1)} \right)
 \right]
\nonumber\\
+
\int dx' \phi_{g/h}(x')
\int dx \left[
  T_{g/A}^{(I)}(x)\, C_A\, \hat{\sigma}_{gg}^{(0)}
+ T_{g/A}^{(F)}(x)\, C_A\, \hat{\sigma}_{gg}^{(0)}
 \right] 
\Bigg)\, ,
\label{nrqcd-qt2-d}
\een 
which has a very similar expression as that in Eq.~(\ref{hq-diff})
derived in CEM in last subsection.

If we use the same model for the quark-gluon and gluon-gluon
correlation functions as that used in CEM calculation in last
subsection, we obtain the heavy quarkonium broadening in NRQCD model
as 
\ben
\Delta\langle q_T^2\rangle_{\rm HQ}^{\rm NRQCD}
=\left(\frac{8\pi^2\alpha_s}{N_c^2-1}\, \lambda^2\,A^{1/3}\right)
\frac{(C_F+C_A)\,\sigma_{q\bar{q}}^{(0)}+2C_A\,\sigma_{gg}^{(0)}
                +\sigma_{q\bar{q}}^{(1)}}
     {\sigma_{q\bar{q}}^{(0)}+\sigma_{gg}^{(0)}}\, ,
\label{nrqcd-qt2}
\een
where the leading order cross sections calculated in NRQCD model are
given by 
\ben
\sigma_{q\bar{q}}^{(0,1)}
&=&
\sum_q\int dx'\, \phi_{\bar{q}/h}(x')
\int dx\, \phi_{q/A}(x)\,
\hat{\sigma}_{q\bar{q}}^{(0,1)} \, ,
\nonumber\\
\sigma_{gg}^{(0)}
&=&
\int dx'\, \phi_{g/h}(x')
\int dx\, \phi_{g/A}(x)\, \hat{\sigma}_{gg}^{(0)}\, ,
\label{nrqcd-xsec} 
\een
with the partonic cross sections given in Eqs.~(\ref{sigma-qq-0}),
(\ref{sigma-qq-1}), and (\ref{sigma-gg-0}), respectively. 

From the transverse momentum broadening in Eq.~(\ref{cem-qt2})
calculated in CEM in last subsection and that in Eq.~(\ref{nrqcd-qt2})
calculated in NRQCD model, it is clear that 
the leading double scattering contribution to the broadening 
calculated in these two models have
the same expression if one neglects the $P$-wave contribution in
NRQCD approach.  Since the $P$-wave contribution is smaller than the
$S$-wave contribution, and the gluon-gluon fusion subprocess
dominates the heavy quarkonium cross section, we expect that 
$\Delta\langle q_T^2\rangle_{\rm HQ}^{\rm NRQCD}
\approx \Delta\langle q_T^2\rangle_{\rm HQ}^{\rm CEM}
\approx (2\,C_A/C_F) \Delta\langle q_T^2\rangle_{\rm DY}$.

In both CEM and NRQCD approach to the production of quarkonia,
$H({}^3S_1)$, we can also calculate the broadening effect
on those quarkonia that were produced from the decay of either excited
or high spin states of produced quarkonia, known as 
the feeddown mechanism of the quarkonium production.  
Since the $q_T^2$-moment in Eq.~(\ref{avg-qt2}) is normalized by the
cross section (the 0$^{\rm th}$-moment), and the rescattering takes
place at the parton-level, we expect that the feed-down mechanism is
not very sensitive to the quarkonium broadening while it is much more  
sensitive to the quarkonium production rate.  We will come back to the 
role of the feeddown mechanism in quarkonium broadening in
Sec.~\ref{num} when we present our numerical results.

\subsection{Transverse momentum broadening in nucleus-nucleus
  collisions} 
\label{broadening-aa}

In this subsection, we extend our calculations of heavy quarkonium's
transverse momentum broadening in hadron-nucleus collisions to the
broadening in nucleus-nucleus collisions.  We discuss the similarities
and differences between the hadron-nucleus and nucleus-nucleus
collisions, and the role of transverse momentum broadening in probing
the properties of the dense and hot QCD matter
created in high energy nucleus-nucleus collisions.

\begin{figure}[hbt]
\centering
\psfig{file=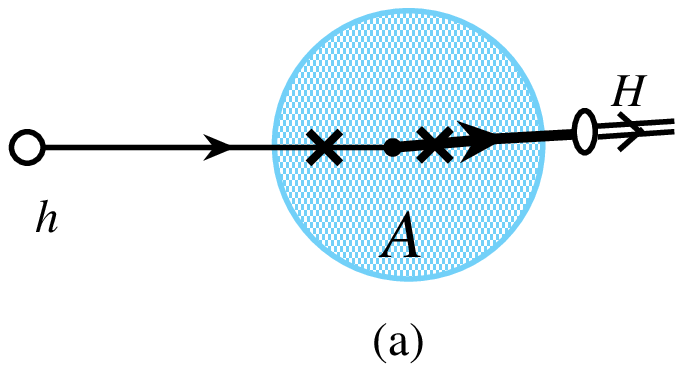,width=2in}
\hskip 0.6in
\psfig{file=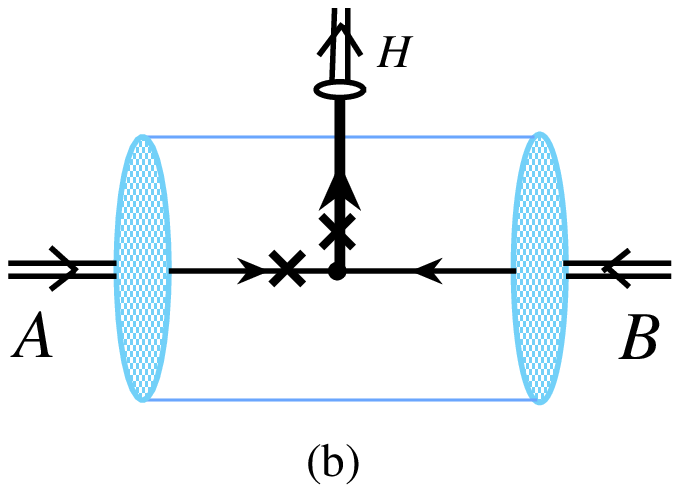,width=2in}
\caption{Sketch of heavy quarkonium production in hadron-nucleus
  collisions as viewed in the target rest frame (a), and that in
  nucleus-nucleus collisions as viewed in the center-of-mass frame
  (b).  The thin and thick lines indicate the incoming parton and the
  outgoing heavy quark pair, respectively.  The cross indicates
  potential rescattering point with soft partons of the nuclear
  medium.}
\label{ha-vs-ab}
\end{figure}

The broadening of heavy quarkonium's transverse momentum is a 
consequence of both initial-state and final-state rescattering in
nuclear medium.  In high energy hadron-nucleus collisions, both
initial-state and final-state rescattering probe the same properties
of a normal nuclear matter.  If the scattering process is viewed in
the rest frame of the nucleus, as sketched in Fig.~\ref{ha-vs-ab}(a),
the incoming active parton and the produced heavy quark pair move very
fast along the direction of the incoming hadron and interact with only
partons of the nucleus near the same impact parameter.  This picture
supports our approximation that the correlation functions for
final-state interaction are about the same as that for initial-state
interaction. If the same process is viewed in the center of mass
frame, we get the same conclusion that only nuclear partons near the
same impact parameter can participate in the rescattering because the
nuclear matter is moving very fast in this frame. 

However, in high energy nucleus-nucleus collisions, the situation can
be very different.  If, other than the hard parton-parton scattering
to produce the heavy quark pair and corresponding rescattering
discussed above, other partons from two colliding nuclei did not
interact in the collision, the transverse momentum broadening of heavy
quarkonium production would be a simple superposition of the
broadening in hadron-nucleus collision.  
In this picture, the leading contribution
to the broadening in a collision between a nucleus $A$ and a nucleus
$B$ would be given by the same expression in Eq.~(\ref{cem-qt2-gg}) 
with the $A^{1/3}$ replaced by $A^{1/3}+B^{1/3}$ or more precisely by 
$L_{AB}/r_0$, where $r_0\approx 0.8$~fm and $L_{AB}$ is an effective
medium length in nucleus-nucleus collision \cite{Miller:2007ri}.
The value of $r_0$ is obtained by letting $L_{pA}\approx
r_0 A^{1/3} \approx (3/4)R_A$ with the nuclear radius $R_A$.

But, as indicated by the early data from RHIC, soft partons from two
nuclei do interact to form a dense and hot QCD quark-gluon medium in
high energy nucleus-nucleus collisions.  
The final-state rescattering between the
produced heavy quark pair of transverse momentum $q_T$ and the almost
stationary or slowly expanding hot QCD medium in the center of mass
frame of nucleus-nucleus collision, as sketched in
Fig.~\ref{ha-vs-ab}(b), is very unlikely to broaden the 
pair's transverse momentum.  Instead, the final-state interaction
could suppress the production rate of the leading (or large momentum) 
colored and coherent heavy quark pair due to the energy loss
\cite{hq-energyloss}, which is responsible for the observed 
jet quenching (or the suppression of leading hadrons or heavy quarks) 
at RHIC \cite{rhic-findings},
and decrease the averaged transverse 
momentum $\langle q_T^2 \rangle$.  On the other hand, 
the initial-state interaction in nucleus-nucleus collisions is likely
to broaden the transverse momentum of the active parton in the same
way as that in hadron-nucleus collisions.  Therefore, the
measured $\langle q_T^2 \rangle$ in nucleus-nucleus collision is 
a consequence of two competing effects: the initial-state interaction
tries to broaden the transverse momentum while the final-state
rescattering in a slowly expanding medium tends to reduce the pair's 
transverse momentum.  The detailed analysis and calculation of the 
competing final-state effects on the quarkonium broadening in 
nucleus-nucleus collisions requires a careful modeling of the 
hot and dense medium and is beyond the scope of this paper.

Precise measurements of transverse momentum broadening of
heavy quarkonium production in relativistic heavy ion collisions
should provide very valuable information on the formation of the
dense and hot quark-gluon medium and its properties.  In
nucleus-nucleus collisions, a deviation of the transverse momentum
broadening from the simple superposition of that measured in
hadron-nucleus collisions clearly indicates a change of nuclear matter
properties from the interaction between soft partons of colliding
nuclei.  It should indicate the formation of a 
dense quark-gluon medium when the measured
transverse momentum broadening is equal or less than the expected
broadening from the initial-state interaction alone.  

If the long-range interaction of soft partons from two colliding
nuclei is quick and strong, and the dense quark-gluon medium
is formed very early in relativistic heavy ion collisions, 
the initial-state interaction 
in nucleus-nucleus collisions could be different from a
superposition of the initial-state effect in hadron-nucleus
collisions due to the modification of nuclear matter.  
In order to independently test the initial-state 
effect from the final-state rescattering, we
calculate the transverse momentum broadening of $Z$ (as well as $W$)
bosons in relativistic heavy ion collisions at the LHC in next
section.

\section{Transverse momentum broadening of $Z$ (and $W$) production at
  the LHC} 
\label{WZ}

The lack of the final-state interaction of a $Z$ (or a $W$) boson 
when it is reconstructed from its leptonic decay channels 
makes its transverse momentum broadening 
in high energy nuclear collisions an ideal probe
for the initial-state interaction, in particular, the density of
nuclear medium in the early stage of relativistic heavy ion
collisions \cite{CERN-Yellow}.  If the long-range soft gluon
interactions between two incoming heavy ions were so strong and a
dense nuclear medium was formed before the short-distance creation of
a $Z$ (or a $W$) boson, the transverse momentum broadening would be a 
clean probe of the density of the dense medium.  Otherwise, the
transverse momentum broadening of $Z$ (or $W$) bosons in
nucleus-nucleus collisions would be a simple superposition of that in
hadron-nucleus collision.  Therefore, by measuring the broadening of
$Z$ (or $W$) bosons in both hadron-nucleus and nucleus-nucleus
collisions, we could learn valuable information on whether the dense
quark-gluon medium could be formed at a very early stage in
relativistic heavy ion collisions \cite{Kang:2007bn}.  

\begin{figure}[hbt]
\centering
\psfig{file=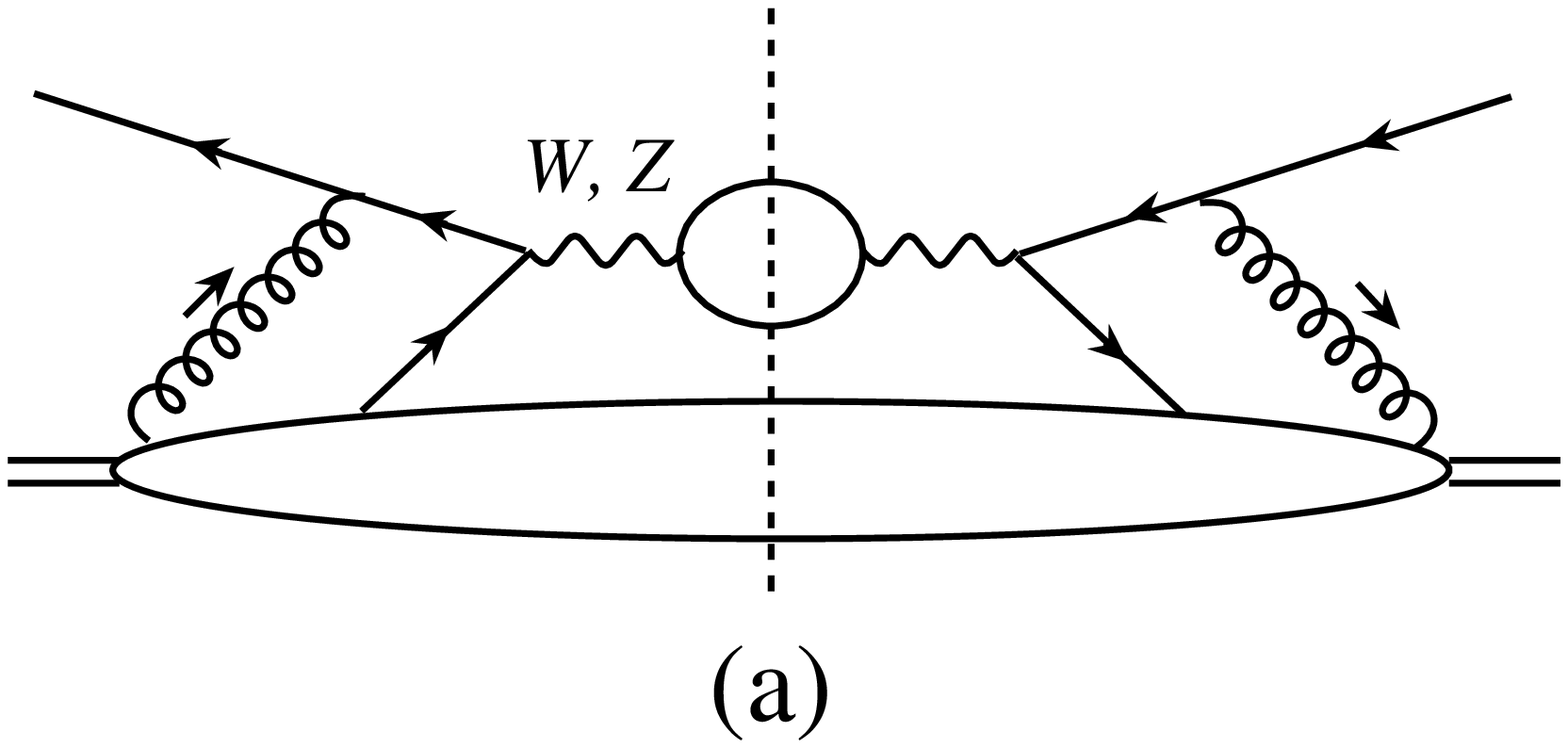,width=2.2in}
\hskip 0.6in
\psfig{file=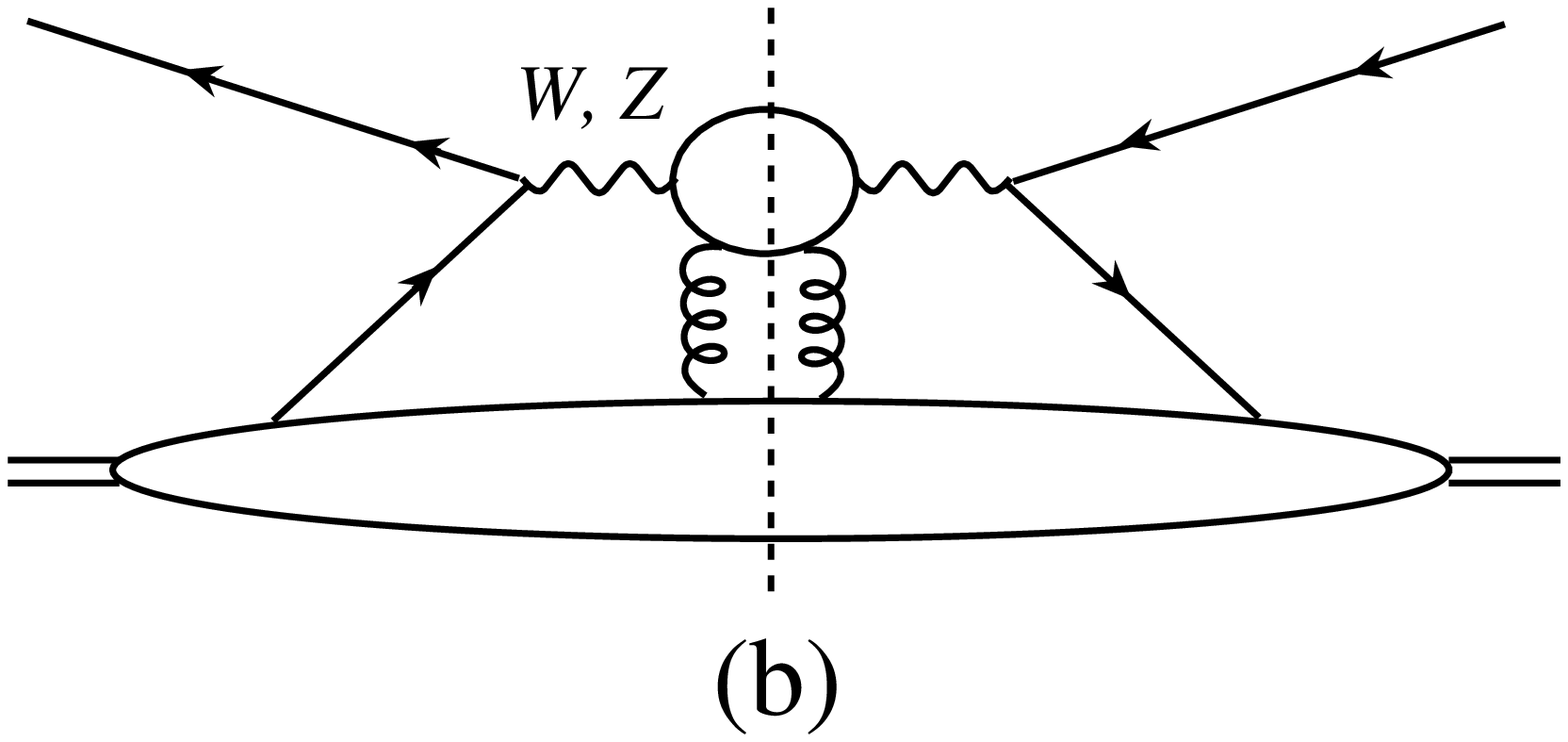,width=2.2in}
\caption{Leading order double scattering diagrams for production of 
  a $Z$ (or a $W$) boson in hadron-nucleus collisions via the
  initial-state interaction (a) and the possible final-state
  rescattering if the vector boson is reconstructed from its hadronic
  decay. }
\label{wz-d}
\end{figure}

In Fig.~\ref{wz-d}, we sketch the leading double scattering diagrams
for the $Z$ (or $W$) production in hadron-nucleus collisions.  The
diagram in Fig.~\ref{wz-d}(a), which is almost identical to that for 
the Drell-Yan transverse momentum broadening, represents the
initial-state interaction, while that in Fig.~\ref{wz-d}(b) represents
the possible final-state rescattering if the vector boson could be
reconstructed from its hadronic decay channels.
For the $Z$ and $W$ bosons reconstructed from their leptonic decay
channels, their transverse momentum broadening is mainly caused by the 
initial-state interaction, just like the broadening of the virtual
photon in the Drell-Yan production.  
From the leading double scattering diagram in Fig.~\ref{wz-d}(a) and 
following the same derivation for the Drell-Yan broadening 
in Sec.~\ref{drellyan}, we obtain the leading transverse
momentum broadening of $Z$ (or $W$) bosons in hadron-nucleus
collisions as, 
\ben
\Delta\langle q_T^2\rangle_{hA}^Z
&=&
C_F\, \frac{8\pi^2\alpha_s(M_Z)}{N_c^2-1}\, \lambda_Z^2\, A^{1/3}\, ,
\nonumber \\
\Delta\langle q_T^2\rangle_{hA}^W
&=&
C_F\, \frac{8\pi^2\alpha_s(M_W)}{N_c^2-1}\, \lambda_W^2\, A^{1/3}\, ,
\label{qt2-wz}
\een
where $\lambda_Z^2$ ($\lambda_W^2$) is the nonperturbative scale for
the double scattering production of $Z$ (or $W$) bosons.  It is
defined in the same was as that in Eq.(\ref{ansatz}) and represents a
ratio of nuclear four parton correlation function over a corresponding
normal parton distribution.  
It is proportional to an averaged gluon field strength
square in nuclear matter, 
$\langle F^{+\alpha}F^{+}_{\ \alpha} \rangle$ \cite{Qiu-Vitev}.
The $\lambda_Z$ should be proportional to the typical
transverse momentum or effective virtuality of soft gluons
participating in the double scattering.  As we will discuss in next
section, the value of the $\lambda^2_Z$ (or $\lambda^2_W$) could
depend on the momentum exchange of the hard collision, $Q\sim M_Z$ (or
$M_W$), as well as the collision energy.

If we assume that the dense quark-gluon medium in relativistic heavy ion 
collisions is not formed before the creation of the heavy vector boson
$Z$ (or $W$), we could apply our formula in Eq.~(\ref{qt2-wz}) 
to the broadening in nucleus-nucleus collisions as a simple
superposition of the hadron-nucleus collision, and obtain the transverse
momentum broadening in nucleus-nucleus collision as 
\ben
\Delta\langle q_T^2 \rangle_{AB}^V
\approx
C_F\, \frac{8\pi^2\alpha_s(M_{V})}{N_c^2-1}\, 
\lambda_{V}^2\, \frac{L_{AB}}{r_0}\, ,
\label{qt2-aa}
\een
where $V=Z,W$ for the $Z$ and $W$ production, respectively.
If the long-range soft gluon interactions between two colliding heavy
ions are so strong that the dense quark-gluon medium was 
formed before the short-distance creation of a $Z$ (or a $W$) boson, we
expect the same formula in Eq.~(\ref{qt2-aa}) to be valid for the
leading contribution to the transverse momentum broadening, but, with
a new $\lambda_V^2$ different from that of a normal nuclear matter.
The value of the effective $\lambda_V^2$ and its dependence on the
collision geometry and collision energy should provide valuable
informations on the formation of the QCD medium and its properties.

\section{Numerical results}
\label{num}

In this section, we provide a numerical comparison between our
calculated heavy quarkonium broadening in nuclear collisions 
with existing data from Fermilab and RHIC experiments, and make
predictions for the transverse momentum broadening at the LHC. 

We have calculated in this paper the transverse momentum
broadening of vector boson production in nuclear collisions in terms
of the QCD factorization approach.  We factorized the rescattering
contribution to the broadening into two parts: (1) the 
non-perturbative, but, well-defined universal parton-parton 
correlation functions, which represent the probability to
find the scattering centers in the nuclear medium, and (2)
corresponding parton-level rescattering subprocess, which are
infrared safe and perturbatively calculable.
As a result of the factorization, the normalization for the transverse
momentum broadening from partonic double scattering is 
uniquely fixed by the size of the non-perturbative 
quark-gluon and gluon-gluon correlation functions.  
If we use the model in Eqs.~(\ref{ansatz}) and (\ref{ansatz-g}) 
to parameterize the correlation functions, 
the numerical results of our calculated transverse momentum broadening
should be directly proportional to the value of the $\lambda^2$. 

The value of the $\lambda^2$, or more precisely,
the value of the parton-parton correlation functions should not depend
on which vector boson was produced.  This is because of the
universality and the long-distance nature of the correlation
functions.  However, the value of the correlation functions or the
$\lambda^2$ should depend on the momentum scale at which the hard part
of the partonic scattering was evaluated.  

As explicitly demonstrated in last three sections, all phase space
integrations for the partonic rescattering can be deformed into
perturbative region, and both initial- and final-state parton-level
rescattering are evaluated at a hard scale $Q\sim 2m_Q$ or $M_Z$ (or
$M_W$) for heavy quarkonium or $Z$ (or $W$) production, respectively.  
From the model in Eq.~(\ref{ansatz}) (or (\ref{ansatz-g})), the
non-perturbative parameter, $\lambda^2$, represents a ratio of nuclear 
four parton correlation function over a normal parton distribution.
As demonstrated in Ref.~\cite{Qiu-Vitev} by approximately decomposing
a nuclear state into a product of nucleon states, the ratio (so as the
$\lambda^2$) can be expressed in terms of an averaged gluon field
strength square, $\langle F^{+\alpha}F^{+}_{\ \alpha} \rangle$.  In
this picture, the $\lambda$ represents the virtuality or the typical
transverse momentum of the partons participating in the partonic
rescattering, and certainly depends on the momentum exchange of the
hard collision, $Q$, as well as the collision energy, $\sqrt{s}$,
which determines the available phase space for the collision.
For hadron-nucleus collisions with a large momentum transfer, $Q$, we
expect the $\lambda^2$ to be proportional to $\ln(Q^2)$  
from the parton shower leading to the hard collision.  
If the collision energy $\sqrt{s}$ is very large and the momentum
fraction of the active parton, $x$, is very small, we would expect the
$\lambda^2$ to be proportional to the saturation scale $Q_s^2 \propto
1/x^\delta$ with $\delta\approx 0.3$ 
\cite{Golec-Biernat:1998js,Iancu:2003xm}.
For the vector boson production, the typical momentum fraction of an
active parton, $x\sim Q/\sqrt{s}$.  Therefore, we expect 
$\lambda^2(Q) \propto \ln(Q^2)\,(\sqrt{s}/Q)^\delta$. 

We use data on the Drell-Yan broadening in hadron-nucleus collisions,
which were measured by Fermilab experiments at the fixed target
energy, $\sqrt{s}=38.8$~GeV \cite{jcpeng}, to extract the
$\lambda_{\rm DY}^2$.  From the value of $\lambda_{\rm DY}^2$, we
estimate the value of $\lambda^2$ for producing a vector boson of
invariant mass $Q$ at a collision energy $\sqrt{s}$ as follows, 
\ben
\lambda^2(Q)\approx
\lambda^2_{\rm DY}\
\frac{\ln(Q^2)}{\ln(\langle Q_{\rm DY}^2\rangle)}\,
\left(\frac{\sqrt{s}/Q}
           {38.8/\langle Q_{\rm DY}\rangle}\right)^{\delta}\, ,
\label{lambda-Q}
\een
with the $\sqrt{s}$ in unit of GeV, $\langle Q_{\rm DY}\rangle \sim
6$~GeV, and $\delta\sim 0.3$.

\begin{figure}[htb]
\centering
\psfig{file=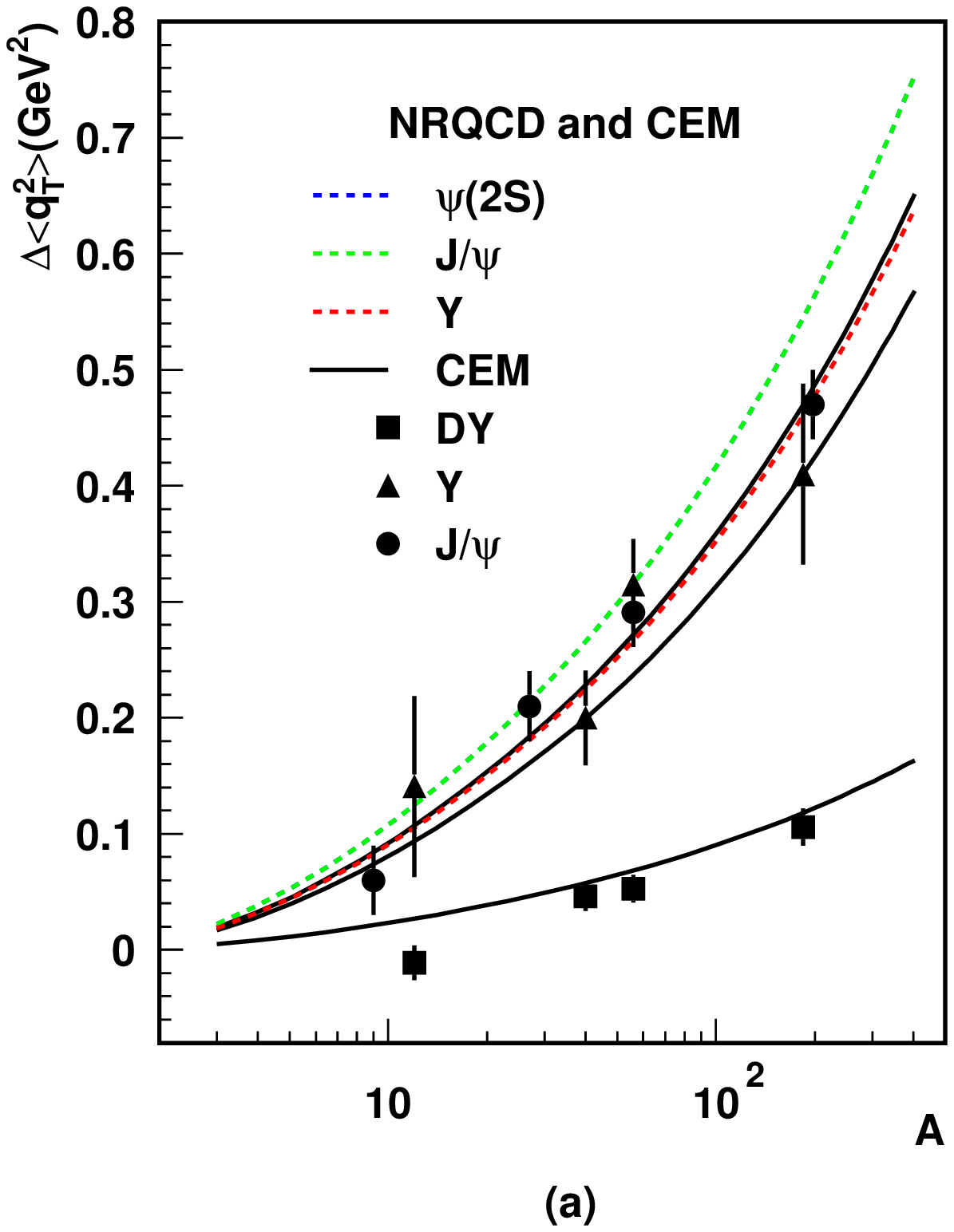, width=2.5in}
\hskip 0.5in
\psfig{file=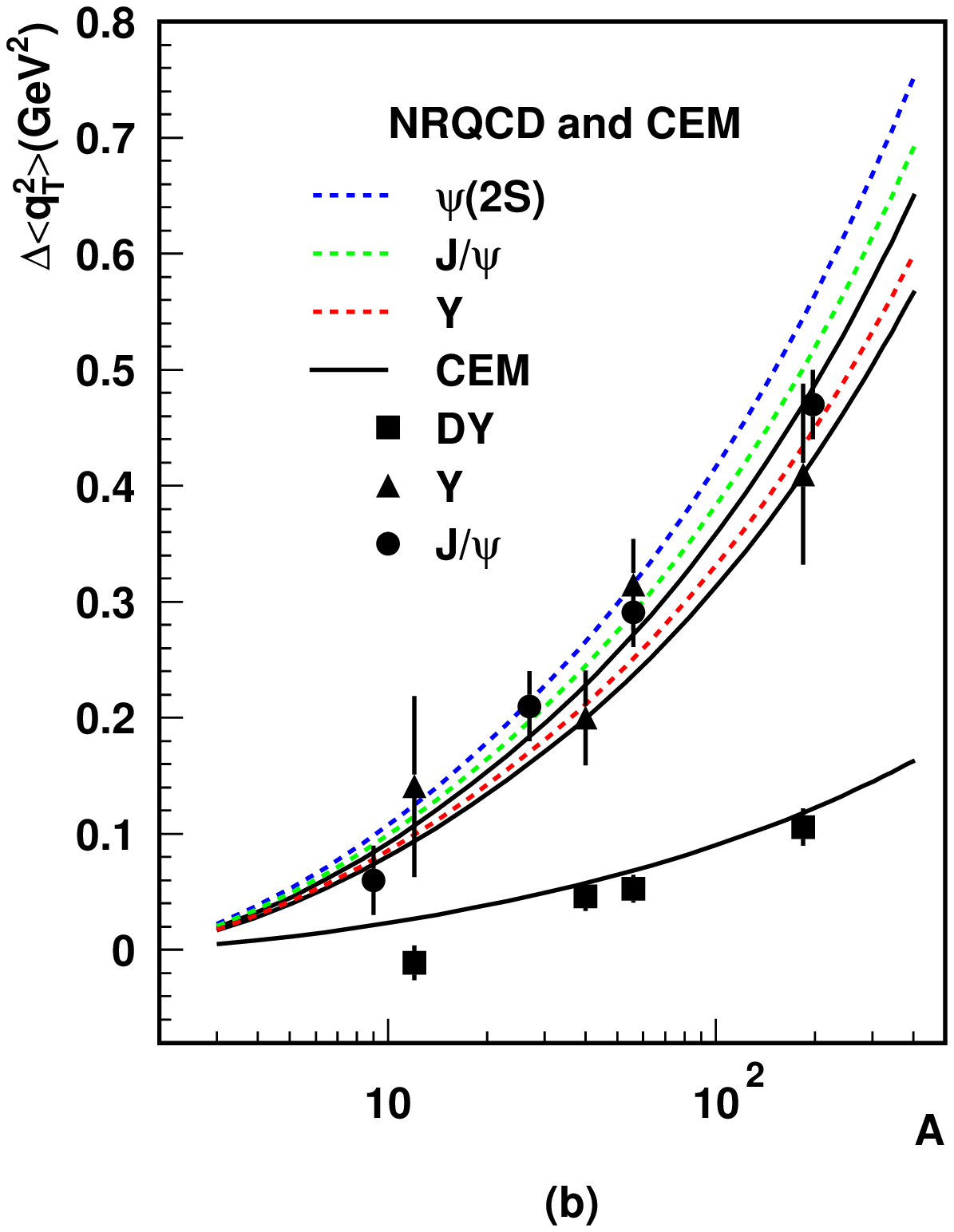, width=2.5in}
\caption{Data on transverse momentum broadening of heavy quarkonium
  as well as Drell-Yan production in hadron-nucleus collisions. 
  Also plotted are theoretical calculations using Eq.~(\ref{cem-qt2})
  (solid lines) and Eq.~(\ref{nrqcd-qt2}) (dashed lines), derived by
  using CEM and NRQCD model, respectively.  Three solid lines (from
  the top to bottom) correspond to $J/\psi$, $\Upsilon$, and
  Drell-Yan, while three dashed lines represent $\psi'$, J/$\psi$, and
  $\Upsilon$ production from NRQCD model.   The quarkonium broadening
  calculated in NRQCD model is evaluated with (a) and without (b)
  quarkonia from the feeddown mechanism.}
\label{fixed-tgt}
\end{figure}

Fermilab experiments: E772, E789, and E866 have measured 
the transverse momentum broadening of the Drell-Yan, as well as 
J/$\psi$, $\psi'$, and $\Upsilon$ production in hadron-nucleus
collisions \cite{jcpeng,fix}.  In Fig.~\ref{fixed-tgt}, we plot the
data on both the Drell-Yan and heavy quarkonium broadening as a function
of atomic weight of nuclear targets.  The broadening for the
data was defined as a difference between the $q_T^2$-moment in
proton-nucleus and proton-deuteron collisions:  
$\Delta\langle q_T^2\rangle=\langle q_T^2 \rangle_{pA}-\langle
q_T^2 \rangle_{pD}$.  By fitting the data on the Drell-Yan broadening
as a function of $A^{1/3}-2^{1/3}$, we obtain 
$\lambda_{\rm DY}^2 \approx 0.01$~GeV$^2$, which 
gives the bottom solid line for the Drell-Yan broadening in
Fig.~\ref{fixed-tgt}, and is consistent with the value extracted in 
Ref.~\cite{xiaofeng}.  

In Fig.~\ref{fixed-tgt}(a), we plot our theoretical calculations of
transverse momentum broadening of {\it direct} heavy quarkonium 
production in hadron-nucleus collisions 
at the Fermilab fixed target energy, $\sqrt{s}=38.8$~GeV.
To obtain the numerical results of theoretical calculations, we 
use CTEQ6L for nucleon parton distribution functions \cite{CTEQ} 
and EKS98 parametrization for nuclear parton distributions (nPDF) 
\cite{EKS98} to evaluate the leading order production cross sections 
in Eqs.~(\ref{cem-qt2}) and (\ref{nrqcd-qt2}).  
The non-perturbative parameter, $\lambda^2$, in Eqs.~(\ref{cem-qt2}) 
and (\ref{nrqcd-qt2}) could be slightly different for J/$\psi$ and
$\Upsilon$ production due to the difference in the scale of hard
collision, $Q\sim 2m_Q$.  
Using $\lambda_{\rm DY}^2 \approx 0.01$~GeV$^2$, 
$M_{{\rm J/}\psi}=3.1$~GeV, $M_{\Upsilon}=9.5$~GeV, we estimate 
from Eq.~(\ref{lambda-Q}) that  
$\lambda_{{\rm J/}\psi}^2 \approx 0.008$~GeV$^2$ and 
$\lambda_{\Upsilon}^2 \approx 0.011$~GeV$^2$ at $\sqrt{s}=38.8$~GeV. 
We use Eq.~(\ref{nrqcd-qt2}) to evaluate the broadening for $\psi'$,
J/$\psi$, and $\Upsilon$ (the dashed lines) in NRQCD model.  The NRQCD
matrix elements are taken from Refs.~\cite{qwg-review,Braaten:2000cm}.  
The small difference between the broadening of J/$\psi$ and
$\Upsilon$ is caused by the relative size between the quark and gluon
contribution due to different sizes of matrix elements and the range
of parton momentum fractions.  For the direct production, J/$\psi$
and $\psi'$ have almost the identical broadening as shown in
Fig.~\ref{fixed-tgt}(a). 
Similarly, we use Eq.~(\ref{cem-qt2}) to evaluate the broadening for
J/$\psi$ (top solid line) and $\Upsilon$ (middle solid line) in CEM
with $m_Q=1.5$~GeV and $M_Q=1.85$~GeV for J/$\psi$ and
$m_Q=4.5$~GeV and $M_Q=5.3$~GeV for $\Upsilon$ production,
respectively.  The transition probability, $F_{Q\bar{Q}\to H}$,
cancels between the numerator and the denominator in
Eq.~(\ref{cem-qt2}).  
The small difference between these two solid lines are again caused by
the relative size of the quark and gluon contribution due to the
slightly different ranges of parton momentum fractions for 
J/$\psi$ and $\Upsilon$ production.  

We also test the effect of transverse momentum broadening on heavy
quarkonia produced by the feeddown mechanism.  Using the partonic hard
parts derived in Sec.~\ref{broadening-nrqcd}, 
we calculate in NRQCD model the transverse momentum broadening of
quarkonia in high spin or excited states, which then 
decay into the observed spin-1 and $S$-wave quarkonia, $H({}^3H_1)$.  
We plot the calculated broadening with this feeddown
mechanism in Fig.~\ref{fixed-tgt}(b).  
Although the feeddown mechanism could provide as much
as 40\% of J/$\psi$ production rate, its net effect on the transverse
momentum broadening is very small because of the fact that the
broadening defined in Eq.~(\ref{avg-qt2}) is normalized by the
inclusive cross section.

For a large nucleus with the atomic weight $A\ge 10^2$, some of the
produced heavy quark pairs could transmute to a color singlet 
pre-quarkonium state (or even a physical quarkonium) before exiting
the nuclear matter.  Comparing to a colored heavy quark pair, these
color singlet states should have a weaker interaction with the nuclear
matter and get less broadening in transverse momentum.  Therefore, 
we expect the theoretical curves in Fig.~\ref{fixed-tgt} to be
slightly less steep than what were shown when $A\ge 10^2$. 

From Fig.~\ref{fixed-tgt}, we conclude that perturbative QCD
calculations of the quarkonium broadening based on both CEM and NRQCD
model give a good description of existing experimental data 
in hadron-nucleus collisions.  
The major difference between the heavy quarkonium and the
Drell-Yan broadening is naturally explained by the role of final-state 
interactions.  Although the production mechanism in CEM and NRQCD
model is different, these two models of heavy quarkonium production
predict almost the same functional form for the transverse momentum
broadening, as shown in Eq.~(\ref{cem-qt2}) and Eq.~(\ref{nrqcd-qt2}),
respectively.  Since the $P$-wave contribution is relatively small,
as shown in Fig.~\ref{fixed-tgt}, these two models predict almost 
the same transverse momentum broadening.  In addition, both models 
predict that J/$\psi$ and $\Upsilon$ have effectively the same  
broadening in hadron-nucleus collisions other than a small difference
caused by the available phase space (i.e., the available range of
parton momentum fractions).


\begin{figure}[htb]
\centering
\psfig{file=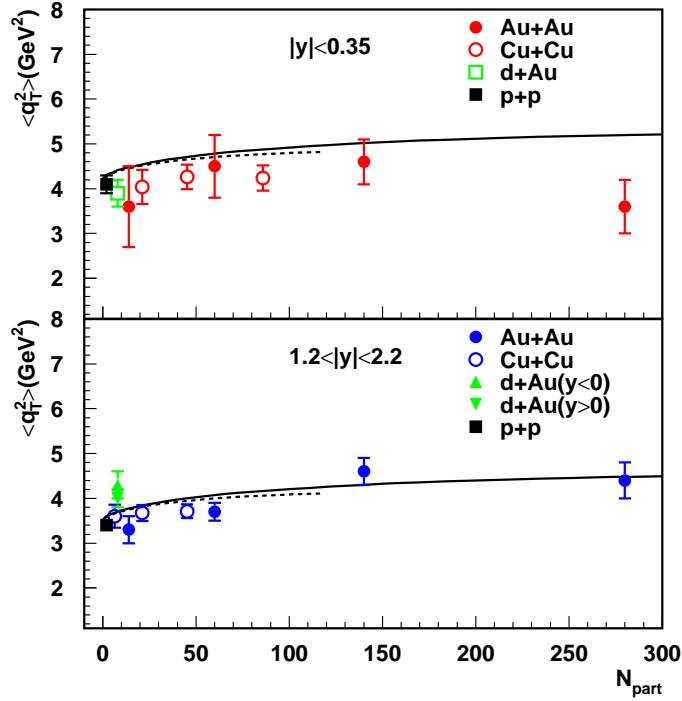, width=3.5in}
\caption{RHIC data on averaged transverse momentum square of J/$\psi$
  production as a function of the number of participants
  \cite{PHENIX-data}.  The top panel is for the J/$\psi$'s produced in
  the central rapidity region while the bottom panel is for those
  produced in more forward or backward region.  Also plotted are
  theoretical calculations using Eq.~(\ref{qt2-rhic}).  Solid lines
  for Au-Au collisions and dashed lines for Cu-Cu collisions,
  respectively. } 
\label{rhic-aa}
\end{figure}

In Fig.~\ref{rhic-aa}, we plot PHENIX data on averaged transverse
momentum square of J/$\psi$ production at RHIC energies as a function
of the number of participants, N$_{\rm part}$ \cite{PHENIX-data}.
The top panel is for the J/$\psi$ produced in the central rapidity
region with $|y|<0.35$, and the bottom is for those produced in the
forward and backward region with $1.2<|y|<2.2$.  
We also plot our theoretical calculations of the transverse momentum 
square by using
\ben
\langle q_T^2 \rangle_{AB} 
\approx \langle q_T^2 \rangle|_{pp-exp} 
+ \Delta \langle q_T^2 \rangle_{AB} \, ,
\label{qt2-rhic}
\een
where $\langle q_T^2 \rangle|_{pp-exp}$ is fixed by the data point
from proton-proton collision in Fig.~\ref{rhic-aa} and 
$\Delta \langle q_T^2 \rangle_{AB}$ is given by our calculation.
We evaluate $\Delta \langle q_T^2 \rangle_{AB}$ in nucleus-nucleus
collisions with an effective medium length $L_{AB}$ as
\ben
\Delta\langle q_T^2\rangle_{AB\to{\rm J/}\psi}^{\rm CEM}
=\left(\frac{8\pi^2\alpha_s}{N_c^2-1}\, 
 \lambda_{{\rm J/}\psi}({\rm RHIC})^2\, \frac{L_{AB}}{r_0} \right) 
\frac{ (C_F+C_A)\, \sigma_{q\bar{q}}+ 2\,C_A\, \sigma_{gg}}
     {\sigma_{q\bar{q}}+\sigma_{gg}}
\label{cem-qt2-ab}
\een
in CEM.  Eq.~(\ref{cem-qt2-ab}) is obtained from
Eq.~(\ref{cem-qt2}) by replacing the $A^{1/3}$ by
$L_{AB}/r_0$.  We can calculate the broadening in NRQCD model by
applying the same replacement to Eq.~(\ref{nrqcd-qt2}).
At the RHIC energy, we obtain
$\lambda_{{\rm J/}\psi}({\rm RHIC})^2 \approx 0.013$~GeV$^2$
from Eq.~(\ref{lambda-Q}).
We calculate the $L_{AB}$ in Glauber model and convert it to 
${\rm N}_{\rm part}$ \cite{Miller:2007ri}.  

In nucleus-nucleus collisions, soft gluons from the colliding ions 
can interact even if the collision is not a head-on or a central
collision.  Such interaction in a non-central collision may not be
strong enough to stop all soft partons to form an almost stationary or
slowly expanding hot medium of quarks and gluons.  It certainly can
slow down some of the colliding soft partons to change the
characteristics of the nuclear matter, which could alter the
final-state interactions. As a result, the final-state interaction
between the produced heavy quark pairs and the modified nuclear matter
in nucleus-nucleus collisions, as sketched in Fig.~\ref{ha-vs-ab}(b),
generates less transverse momentum broadening 
if it does not reduce the transverse momentum due to 
energy loss.  In Fig.~\ref{rhic-aa}, we plot our calculations of
J/$\psi$ transverse momentum broadening by using
Eqs.~(\ref{qt2-rhic}) and (\ref{cem-qt2-ab}), and keeping only the
contribution from initial-state rescattering.  
The solid lines are for the Au-Au collision, while the dashed lines
are for the Cu-Cu collision.  Our calculations are consistent with the
data in both rapidity regions.  

In the central Au-Au collision, a hot and dense medium is produced.  
As discussed above, the averaged transverse momentum could be reduced,
instead of the broadening, due to the energy loss of the produced
heavy quark pairs when they interact with the hot and slowly expanding
medium.  A more detailed study of the momentum shift of the heavy
quark pairs in such a hot medium could provide a more accurate
description of the data in central region, and help the extraction of
medium properties.


Transverse momentum broadening of $Z$ (or $W$) bosons in high energy
nuclear collisions could provide a clean measurement of initial-state
interactions, and help isolating final-state rescattering effect in
heavy quarkonium production.  However, because of the heavy mass of 
$Z$ (or $W$) boson, only the LHC has a chance to measure the
broadening reliably \cite{Kang:2007bn}.  

\begin{figure}[hbt]
\centering
\psfig{file=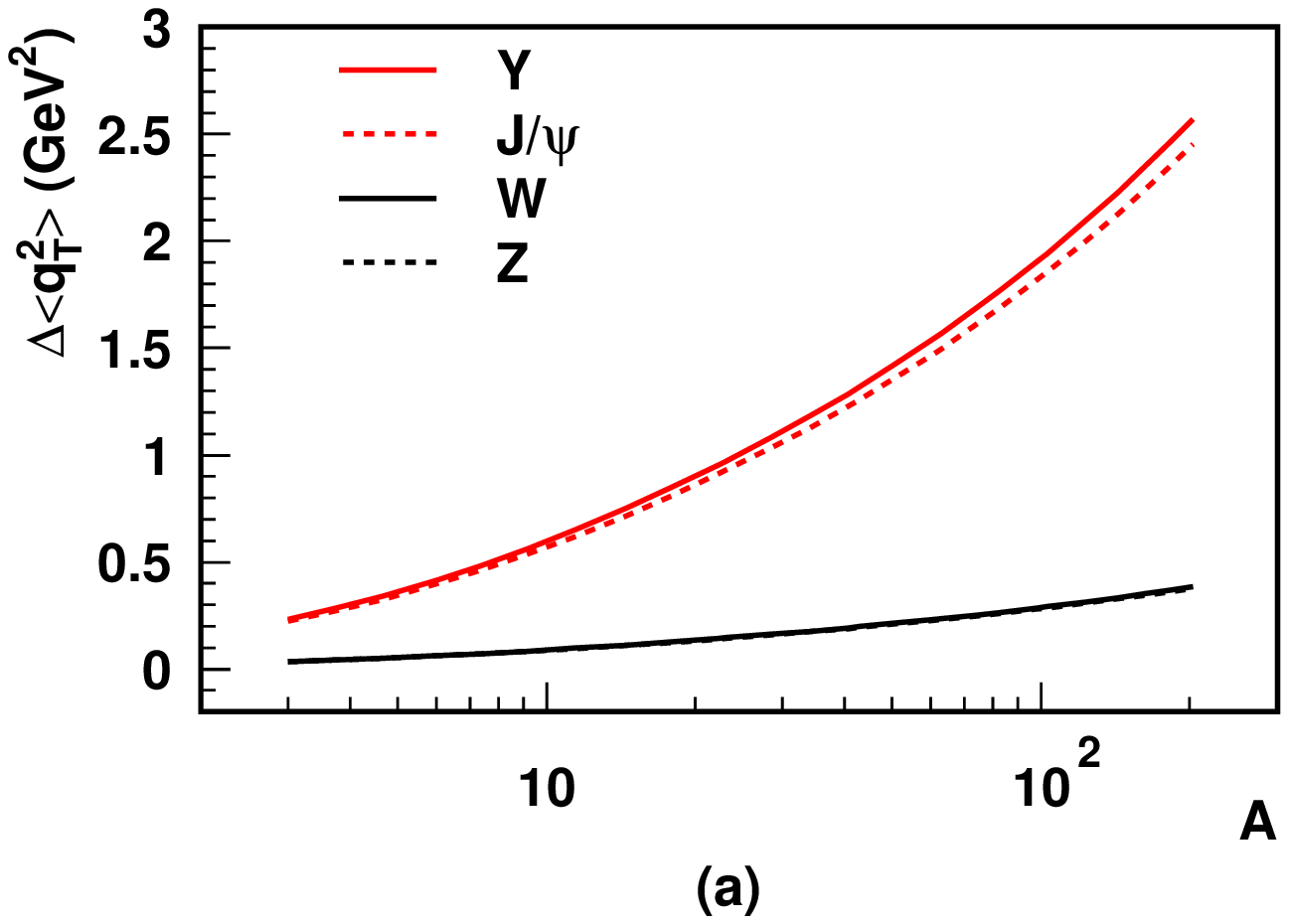,height=2in}
\hskip 0.4in
\psfig{file=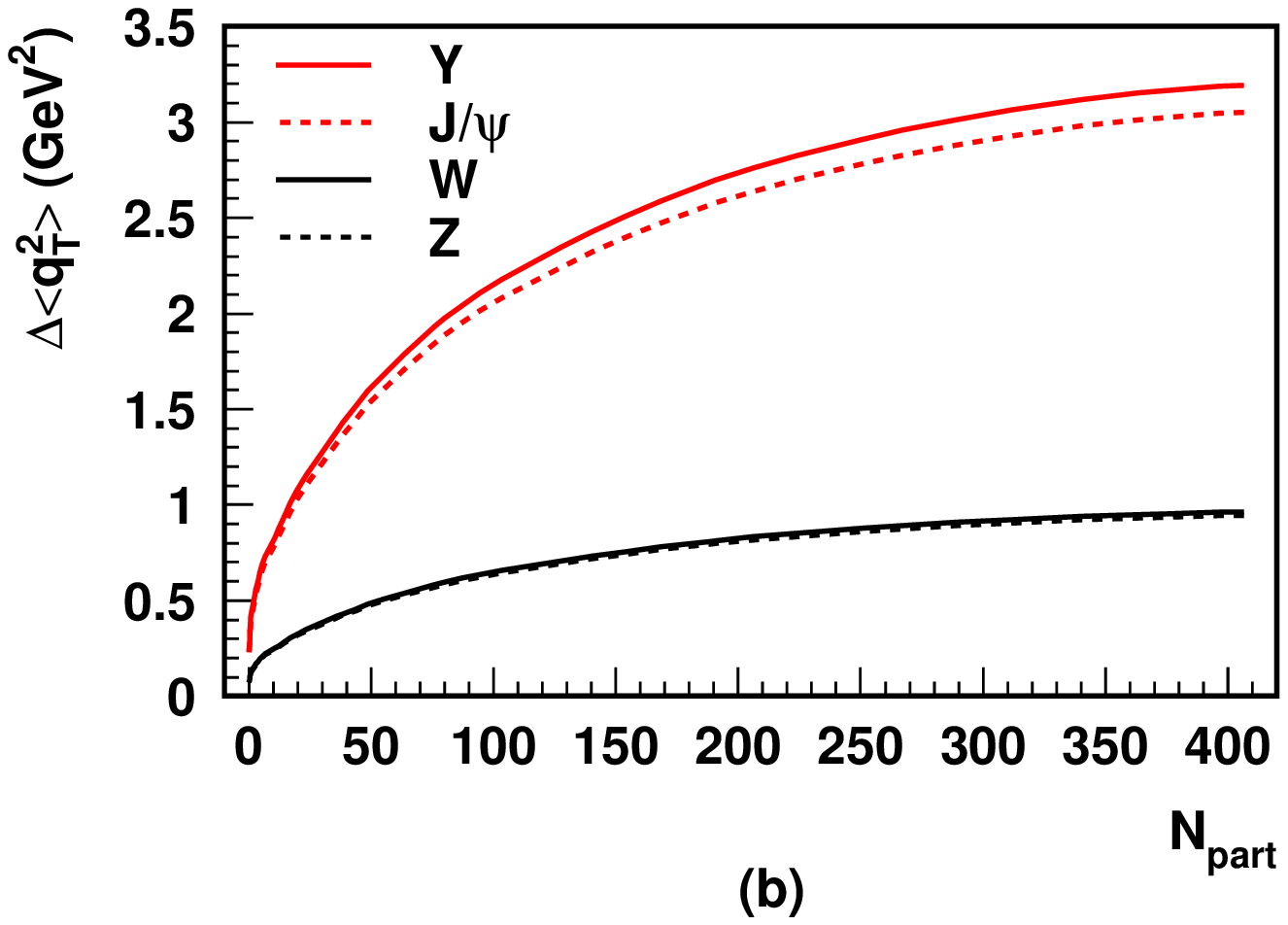,height=2in}
\caption{Transverse momentum broadening of $Z$ and $W$
  (lower set of curves) in hadron-nucleus collisions (a) and
  nucleus-nucleus collisions (b) at $\sqrt{s}=5.5$~TeV as a
  function of atomic weight of nuclear beam and the number of
  participants, ${\rm N}_{\rm part}$, respectively. 
  Also plotted are predictions (upper set of curves) 
  for the transverse momentum broadening of heavy quarkonium 
  production in hadron-nucleus collision at the
  LHC energy (a); and the initial-state only broadening in
  nucleus-nucleus collision at the same energy (b). } 
\label{wzbroaden}
\end{figure}

We use Eq.~(\ref{lambda-Q}) to estimate the $\lambda^2$ for the heavy
vector boson production at $\sqrt{s}=5.5$~TeV, the averaged
nucleon-nucleon collision energy in relativistic heavy ion collisions
at the LHC.  We obtain $\lambda^2({\rm LHC}) \approx 0.035, 0.05$ and 
$0.05$~GeV$^2$ for the production of J/$\psi$, $\Upsilon$, and the
heavy vector boson $Z$ (or $W$), respectively.  Although the
$\Upsilon$ mass is much smaller than that of a $Z$ boson, the 
$\Upsilon$ and $Z$ have the same $\lambda^2({\rm LHC})$ for the
transverse momentum broadening due to the larger available phase space
for $\Upsilon$ production \cite{Berger:2004cc}.    
In Fig.~\ref{wzbroaden}, we present our predictions for the transverse
momentum broadening of vector boson production at the LHC.
Using the estimated $\lambda^2_{Z/W}({\rm LHC})\approx 0.05$~GeV$^2$
and Eq.~(\ref{qt2-wz}), we evaluate the transverse momentum broadening
of $Z$ (and $W$) bosons reconstructed from their leptonic decays, and
plot the predictions for hadron-nucleus collisions as a function of
atomic weight of the nucleus, $A$, in Fig.~\ref{wzbroaden}(a).  
We also plot the expected transverse momentum broadening of J/$\psi$
and $\Upsilon$ production in hadron-nucleus collisions in
Fig.~\ref{wzbroaden}(a).  The curves for heavy quarkonium broadening
are evaluated by using 
$\lambda^2_{{\rm J/}\psi}({\rm LHC})\approx 0.035$~GeV$^2$ 
and $\lambda^2_{\Upsilon}({\rm LHC})\approx 0.05$~GeV$^2$, and 
Eq.~(\ref{cem-qt2}) from CEM without contributions from the feeddown
mechanism.  Eq.~(\ref{nrqcd-qt2}) derived from NRQCD model gives the
similar results.  The heavy quarkonium broadening in
Fig.~\ref{wzbroaden}(a) is much larger than that of $Z$ (or $W$) 
bosons because of the additional final-state effect, 
and the difference in color factor and the strength of the
strong coupling constant, $\alpha_s(Q)$. 

In Fig.~\ref{wzbroaden}(b), we plot the expected transverse momentum
broadening of vector boson production in Pb-Pb collision at
$\sqrt{s}=5.5$~TeV at the LHC as a function of the number of
participants, N$_{\rm part}$ \cite{Miller:2007ri}.
We calculate the $L_{AB}$ in Glauber model with inelastic 
nucleon-nucleon cross section $\sigma_{NN}^{in}=70$mb at the 
LHC energy and convert it to ${\rm N}_{\rm part}$ in Pb-Pb collisions
\cite{Miller:2007ri}.  For the $Z$ (and $W$) broadening, we use 
$\lambda_{Z/W}({\rm LHC})^2\approx 0.05$~GeV$^2$, the same value 
used for the broadening in hadron-nucleus collisions in
Fig.~\ref{wzbroaden}(a).  Since the transverse momentum broadening is
directly proportional to the $\lambda^2$, which is proportional to the
gluon strength in the medium, a deviation from the predicted curves in
Fig.~\ref{wzbroaden}(b) signals the formation of the hot quark-gluon
medium before the creation of the heavy $Z$ (or $W$) bosons.

For J/$\psi$ and $\Upsilon$ production in Fig.~\ref{wzbroaden}(b), we 
use $\lambda^2_{{\rm J/}\psi}({\rm LHC})\approx 0.035$~GeV$^2$ 
and $\lambda^2_{\Upsilon}({\rm LHC})\approx 0.05$~GeV$^2$, and 
Eq.~(\ref{cem-qt2}) from CEM to evaluate the quarkonium broadening.
Eq.~(\ref{nrqcd-qt2}) from NRQCD model gives similar predictions. 
The plotted curves for J/$\psi$ and $\Upsilon$ production in
Fig.~\ref{wzbroaden}(b) include only initial-state interaction.  
As discussed in Sec.~\ref{broadening-aa}, the  
final-state rescattering in nucleus-nucleus collisions is more likely
to reduce the transverse momentum broadening due to the energy loss,
if a slowly expanding medium was produced.  Therefore, we expect the
curves in Fig.~\ref{wzbroaden}(b) to represent the maximum broadening
of J/$\psi$ and $\Upsilon$ that will be seen in relativistic heavy ion
collisions at the LHC \cite{Kang:2007bn}. 

If we could reconstruct $Z$ and $W$ bosons from their hadronic decay
(e.g., into two jets), which might be impossible to do in the LHC
environment, the hadronic jets from $Z$ and $W$ decay will have to
interact with the nuclear medium.  The final-state multiple scattering
could generate momentum imbalance between these two jets and
effectively introduce an apparent mass shift for the $Z$ and $W$
boson \cite{Luo:1993ui}. 
Such mass shift would provide additional information on the
properties of the hot quark-gluon medium.

\section{Summary}
\label{summary}

In this paper we evaluated transverse momentum broadening of
vector boson production in hadron-nucleus and nucleus-nucleus 
collisions in terms of parton-level multiple scattering.  We argued 
that the broadening, $\Delta\langle q_T^2\rangle_{AB} \equiv 
\langle q_T^2\rangle_{AB} - \langle q_T^2\rangle_{hh}$,  
defined as a difference between the averaged transverse momentum
square measured in nuclear collisions and that in hadron-hadron
collisions, can be systematically calculated in perturbative QCD in
terms of the collinear factorization approach \cite{Qiu-Sterman}.
We factorize the contribution to transverse momentum broadening into 
the calculable short-distance partonic rescattering multiplied by
universal parton-parton correlation functions, which specify the
probabilities to find coherent scattering centers in a nuclear medium.   
We explicitly show that the partonic rescattering diagrams
are evaluated at a perturbative hard scale $Q\sim 2m_Q$.  
We verify the universality of the non-perturbative parton-parton
correlation functions by fitting the data on both the Drell-Yan
broadening and the broadening of J/$\psi$ and $\Upsilon$ production to
clearly demonstrate the predictive power of the QCD factorization
approach.  

For the Drell-Yan virtual photon and $Z$ (or $W$) boson production, we  
evaluated transverse momentum broadening from purely initial-state
multiple scattering.  We discussed the scale dependence of the
non-perturbative parameter, $\lambda^2$, and extrapolate it from its
value at the fixed-target energy to the collider energies.  We
calculated the broadening of $Z$ (and $W$) bosons in both
hadron-nucleus and nucleus-nucleus collisions at the LHC.  We
evaluated the $Z$ (and $W$) boson broadening in nucleus-nucleus
collisions as a superpostion of its broadening in hadron-nucleus
collisions, and argued that a deviation from our calculation is a
clear signal indicating that strong interactions between soft partons
of colliding heavy ions took place before the short-distance creation
of $Z$ bosons.  That is, the transverse momentum broadening of $Z$ (or
$W$) bosons could be a clean and excellent probe of the early stage
dynamics of relativistic heavy ion collisions. 

For J/$\psi$ and $\Upsilon$ production, we demonstated that the
final-state interaction between the produced heavy quark pair and the
nuclear medium is very important in understanding the existing data.
We calculated the broadening in both CEM and NRQCD model, and   
clearly showed that the two models generate a small
difference in the broadening and the broadening has a weak dependence
on the feeddown mechanism.  That is, our results are insensitive to
the details of the hadronization mechanism and perturbatively
reliable.  We found that the leading contribution to heavy quarkonium
broadening in hadron-nucleus collisions is about $2C_A/C_F=4.5$ times
the corresponding Drell-Yan broadening, which gives a good description
of the existing Fermilab data.   

We argued that the role of the final-state interaction to the
transverse momentum broadening in nucleus-nucleus collisions could 
be very different from that in hadron-nucleus collisions.  In
hadron-nucleus collisions, both initial-state and final-state
interactions involve only soft partons of the nucleus near the same
impact parameter, and therefore, provide similar contributions
to the transverse momentum broadening as long as the heavy quark pair
stays in a colored state.  On the other hand, soft partons from two
colliding nuclei could strongly interact to produce a slowly expanding  
quark-gluon medium.  This new medium could interact with the produced 
heavy quark pair if the pair is in a color octet state, and the
interaction could be very weak or vanish if the pair is in a singlet
state.  The final-state interaction with this new medium in
nucleus-nucleus collision, as sketched in Fig.~\ref{ha-vs-ab} could
reduce (instead, to broaden) the pair's transverse momentum.   

We predicted the transverse momentum broadening of vector boson 
(J/$\psi$, $\Upsilon$, and $W/Z$) production in hadron-nucleus
collisions and a maximum broadening in nucleus-nucleus collisions at
the LHC in Fig.~\ref{wzbroaden}. 
A more detailed calculation of final-state interaction between the
produced heavy quark pair and the dense and slowly expanding
quark-gluon medium produced in nucleus-nucleus collisions could
provide a better conncetion between the measured average of the
transverse momentum square and the properties of the dense quark-gluon
medium in relativistic heavy ion collisions \cite{Kang-Qiu}.  
Having a precise transverse momentum broadening could also shed some
lights on nonperturbative dynamics that determines the transverse
momentum distribution of vector boson production in low $q_T$ region
\cite{Guo:1999wy}. 

Finally, we emphasize that our QCD factorization approach for
calculating the heavy quarkonium broadening can be applied for
studying the nuclear dependence of quarkonium cross sections in high
energy nuclear collisions \cite{Qiu:1998rz}, as well as the nuclear
dependence of the quarkonium's rapidity and transverse momentum  
distributions if the transverse momentum $q_T$ is large enough
\cite{Luo:1992fz,Qiu-Sterman}.  The advantage of the QCD factorization  
formalism is that we may readily quantify corrections in powers of 
$\alpha_s$. 

\section*{Acknowledgments}

We thank George Sterman for useful discussions on heavy quarkonium
production.  We thank M.J. Leitch and J.C. Peng for helpful 
correspondence on E772 and E866 data. 
J.~Q. thanks the Theory Group of High Energy Physics
Division at Argonne National Laboratory for its hospitality and
support when a part of this work was completed. 
This work was supported in part by the U.S. Department of Energy under
Grants No.~DE-FG02-87ER40371 and Contract DE-AC02-06CH11357, and in
part by the Argonne University of Chicago Joint Theory Institute (JTI) 
Grant 03921-07-137.


\end{document}